\newif\ifuseprd
\newif\ifeprint
\newif\ifdatelast

\useprdtrue
\eprinttrue
\datelastfalse

\documentclass[preprint,
tightenlines,%
floats,prd,eqsecnum,nobibnotes%
,nofootinbib,aps,12pt]{revtex4}
\usepackage{amsmath,amssymb}
\usepackage{graphicx}
\let\oldappendix\appendix
\renewcommand\appendix{\oldappendix%
    \renewcommand\theequation{\thesection.\arabic{equation}}}

\allowdisplaybreaks[3]

\newif\iftoomuchdetail
\toomuchdetailtrue
\toomuchdetailfalse
\newcounter{saveequation}
\newcounter{detailnum}
\newcommand\savetheequation{\theequation}
\newcommand\detailtheequation{%
	  $\delta$\Roman{detailnum}:\roman{equation}}
\newenvironment{detail}{\iftoomuchdetail\sf
         \setcounter{saveequation}{\value{equation}}%
         \setcounter{equation}{0}\addtocounter{detailnum}{1}%
         \renewcommand\theequation\detailtheequation%
         \fi}{
     \iftoomuchdetail%
     \ifnum\value{equation}=0\addtocounter{detailnum}{-1}\fi%
     \setcounter{equation}{\value{saveequation}}%
     \renewcommand\theequation\savetheequation%
     \fi%
     }

\newcommand\abs[1]{\ensuremath{\left\lvert{#1}\right\rvert}}
\newcommand\anti[2]{\ensuremath{\left\{{#1},{#2}\right\}}}
\newcommand\com[2]{\ensuremath{\left[{#1},{#2}\right]}}
\DeclareMathOperator{\Tr}{Tr}
\newcommand\mathone{{\rlap{\kern .25em l}1}}
\newcommand\one{{\ifmmode{\text{\mathone}}\else{\mathone}\fi}}

\newcommand\p{\ensuremath{\partial}}
\newcommand\cD{{\ensuremath{{\mathcal{D}}}}} 
\newcommand\transpose{{\ensuremath{\text{\sf T}}}}
\newcommand\order[1]{{\ensuremath{{\mathcal O}\left({#1}\right)}}}
\newcommand\field[1]{{\ensuremath{\mathbb{{#1}}}}}
\newcommand\ZZ{{\field{Z}}}

\newcommand\ZR{{\field{R}}}

\newcommand\ppwave{{\em pp}~wave}
\newcommand\ppWave{{\em pp}~Wave}

\makeatletter
\def\@strike{\relax\leavevmode
  \ifmmode
    \expandafter\mathpalette\expandafter\math@strike
  \else
    \expandafter\make@strike
  \fi}
\def\math@strike#1#2{%
  \setbox\z@\hbox{$\m@th#1{#2}$}\fin@strike}
\def\make@strike#1{%
  \setbox\z@\hbox{\color@begingroup#1\color@endgroup}\fin@strike}
\def\fin@strike{%
  \@tempdima\dp\z@
  \@tempdimb\ht\z@
  \lower\@tempdima\hbox{\strike@start}%
  \box\z@
  \raise\@tempdimb\hbox{\strike@end}}
\def\strike@start{\special{ps: %
    currentpoint /starty exch def /startx exch def}}
\def\strike@end{
}
\newcommand\fs{\protect\@strike}

\newcommand\citejournal[4]{{\ifuseprd\else\begingroup\em\fi {#4}%
     \ifuseprd\else\endgroup\fi {\bf {#1}}\ifdatelast, {#3} ({#2})\else%
     \ ({#2}) {#3}\fi}}

\providecommand\plb[3]{{\citejournal{#1}{#2}{#3}{Phys.\ Lett.\ B }}}
\providecommand\npb[3]{{\citejournal{{\ifuseprd\bf\else\em\fi B\/}#1}{#2}{#3}{Nucl.\ Phys.\ }}}
\providecommand\jhep[3]{{\citejournal{#1}{#2}{#3}{J.\ High Energy Phys.\ }}}

\providecommand\cqg[3]{{\citejournal{#1}{#2}{#3}{Class.\ Quant.\ Grav.\ }}}
\providecommand\mpla[3]{{\citejournal{{\em A\/}#1}{#2}{#3}{Mod.\ Phys.\ %
   Lett.\ }}}

\providecommand\citeprd[3]{{\citejournal{#1}{#2}{#3}{Phys.\ Rev.\ D }}}

\providecommand\citeprl[3]{{\citejournal{#1}{#2}{#3}{Phys.\ Rev.\ Lett.\ }}}

\providecommand\hepth[1]{{\ifuseprd{\eprint{{\ifeprint\tt\fi hep-th/#1}}}%
                \else{\tt hep-th/{#1}}\fi}}

\newcommand\phepth[1]{{\ifuseprd\else\tt\fi [\hepth{#1}]}}
\newcommand\ct[1]{{\ifeprint\ifuseprd{\em{#1}},\else{\sf {#1}},\fi\fi}}
\newcommand\bt[1]{{\em {#1}},}

\newcommand\ben{\begin{equation}}
\newcommand\een{\end{equation}}
\newcommand\bea{\begin{eqnarray}}
\newcommand\eea{\end{eqnarray}}

\newcommand\tM{\tilde{M}}
\newcommand\tvphi{{\tilde{\varphi}}}
\newcommand\vx{{\vec x}}
\newcommand\bg{{\bar g}}

\newcommand\nn{\nonumber}

\newcommand\Rho{{\mathcal R}}

\newcommand\eg{{\em e.g.\/}}
\newcommand\ie{{\em i.e.\/}}
\newcommand\cf{{\em cf.\/}}

\newcommand\etal{{\em et.\ al.\/}}

\begin{document}

\ifeprint
\setlength{\baselineskip}{1.2\baselineskip} 
\fi

\title{\vspace*{\fill}{\LARGE  \ppWave\ Big Bangs: \\ 
\vspace*{.3\baselineskip} 
\Large Matrix Strings and Shrinking Fuzzy Spheres
}}
\author{Sumit R. Das}\email{das@pa,uky,edu}
\author{Jeremy Michelson}\email{jeremy@pa,uky,edu}
\affiliation{Department of Physics and Astronomy \\
       University of Kentucky \\
       Lexington, KY \ 40506 \\ U.S.A.
\vspace*{3\baselineskip} 
}

\begin{abstract}
\vspace*{\baselineskip}
We find \ppwave\ solutions in string theory with null-like linear
dilatons.  These provide toy models of big bang cosmologies.
We formulate Matrix String Theory in these backgrounds. Near
the big bang ``singularity'', the string
theory becomes strongly coupled but the Yang-Mills description of
the matrix string is {\em weakly} coupled. The presence of a second
length scale allows us to focus on a specific class of non-abelian
configurations, {\em viz.\/} fuzzy cylinders, for a suitable regime of
parameters.
We show that, for a class of \ppwave{s},
fuzzy cylinders which start out big at early times 
dynamically shrink into usual strings
at sufficiently late times.
\vspace*{\fill}
\end{abstract}

\preprint{\parbox[t]{10em}{\begin{flushright}
UK/05-08 \\ {\tt hep-th/0508068}\end{flushright}}}

\maketitle

\ifeprint
\tableofcontents
\fi

\section{Introduction}\label{intro}

In a recent paper, Craps \etal\ \cite{Craps:2005wd} proposed
a string theoretic toy model of big bang cosmology. The background
is the usual flat background of the Type IIA critical superstring with 
string coupling $g_s$ and string length $l_s$, living on a
compact null direction $x^-$ with radius $R$
\begin{equation}
ds^2 = 2 dx^+dx^- + d\vx \cdot d\vx,
\end{equation}
and a dilaton linearly proportional
to the other null direction $x^+$
\begin{equation}
\Phi = -Q x^+,
\label{nulldilaton}
\end{equation}
so that the effective string coupling is
\begin{equation}
\bg_s = g_s e^{-Qx^+}.
\label{bargs}
\end{equation}
The effective string coupling
is small for $x^+ \rightarrow \infty$ for $Q > 0$ and one has the usual
perturbative spectrum, while
for $x^+ \rightarrow -\infty$ the string theory becomes strongly coupled
and the corresponding Einstein metric has a (null-like) big bang singularity.
For $Q < 0$ we have a time-reversed situation where the big bang is
replaced by the 
big crunch. In this paper we will exclusively deal with $Q > 0$.
The main observation of \cite{Craps:2005wd} is that in the big bang
region, the Matrix String
description~\cite{Motl:1997th,Banks:1996my,Dijkgraaf:1997vv}
of the theory 
as a $1+1$ dimensional $U(N)$ 
Yang Mills theory becomes {\em weakly coupled}. 

Let us briefly review the Matrix String Theory setup.
In the absence of
a linear dilaton, the IIA string theory is dual to a IIB string
theory with a string coupling ${\tilde g}_s$ and string length
${\tilde l}_s$ given by
\begin{equation}
{\tilde g}_s = \frac{R}{g_s l_s}, \qquad {\tilde l}_s^2 =\frac{g_s l_s^3}{R};
\end{equation}
living on a compact space-like circle of radius ${\tilde R}$ given by
\begin{equation}
{\tilde R} = \frac{l_s^2}{R};
\end{equation}
and carrying $N$ units of $D1$-brane charge. A non-perturbative
formulation of the theory is therefore given by a Yang-Mills theory
with a dimensional coupling constant 
\begin{equation}
g_{YM} = \frac{R}{g_s l_s^2}.
\end{equation}
For small $g_s$, only the Yang-Mills fields belonging to the Cartan
subalgebra survive, and the theory describes the original
Type IIA string in a light-cone gauge where $x^+$ is identified
with the time $\tau$ of the YM theory.

In the presence of a linear dilaton, it is argued in
\cite{Craps:2005wd} that the Matrix String Theory may be
obtained by simply replacing $g_s \rightarrow \bg_s$ so
that the resulting Yang-Mills theory has a time-dependent coupling
which becomes {\em weak} as $x^+ \rightarrow -\infty$.
One can equivalently write the theory with
a {\em time-independent\/} coupling but on a
$1+1$ dimensional Milne universe~\cite{hs,bhkn,cc,s,lms,lms2,tt,ckr,fm,d,cm,r,%
grs,pb,dp,bdpr}
with a compact spatial direction $\sigma$,
\begin{equation}
ds^2 = e^{2Q\tau}[-d\tau^2 + d\sigma^2] \equiv g_{\alpha\beta}
dx^\alpha dx^\beta.
\label{conmetric}
\end{equation}
This models a big bang singularity at $\tau = -\infty$.
Since the
Yang-Mills theory is weakly coupled at the big bang, all the
non-abelian degrees of freedom are important. The key
point about this model is that while the ``time'' of the Yang-Mills
theory goes over the full range, the space-time as perceived by
the bulk string theory with a time-independent coupling 
has a null-like singularity.

The situation is quite similar to certain time-dependent
backgrounds recently studied in 
two dimensional non-critical string theory~\cite{Das:2004aq,Das:2005jp}.
In these latter cases, the fundamental
formulation is in terms of Matrix Quantum Mechanics while the closed
string theory is described, for example, by collective field theory with space
arising out of the eigenvalues. For the backgrounds of
\cite{Das:2004aq,Das:2005jp} the space-time perceived by closed
strings are geodesically incomplete with a {\em space-like} boundary
even though the time of Matrix Quantum Mechanics goes over the full
range---provided we use a conformal frame in which the coupling
constant of the collective field theory is time-independent. In
terms of the string theory there is a space-like condensation of the
tachyon. A similar phenomenon occurs in several situations in
perturbative string theory, as argued in \cite{McGreevy:2005ci}.

Two dimensional non-critical string theory is in a sense solvable.
The fundamental degrees of freedom are free fermions and one can use
this fact to understand details of what happens near space-like
boundaries. Matrix String Theory for critical strings is not as well
understood. While it is clear that near the big bang singularity
non-abelian degrees of freedom become important, it would be useful to
have an idea of how these excitations disappear at late times. This
motivated us to look for backgrounds where more is known about
non-abelian excitations of matrix theory. 

Typically one expects that a certain class of non-abelian excitations
of Matrix Theory will play an important role in string or M-theory
backgrounds containing fluxes. One example is the fuzzy spheres formed by
the dielectric effect \cite{Myers:1999ps,TV}. However for most backgrounds, the
precise form of the Matrix Theory is not known. Perhaps the most
interesting class of backgrounds for which the Matrix Theory is known
is provided by M-theory \ppwave{s} \cite{Berenstein:2002jq,clp2,ni}. For
{\em constant} dilatons, this matrix theory has been studied extensively~%
\cite{Dasgupta:2002hx,Dasgupta:2002ru,Kim:2002if,Kim:2002zg,jhp,%
Michelson:2004fh}.
Furthermore, using the compactification
procedure of~\cite{Michelson:2002wa}, Matrix String Theory has been
constructed and studied~%
\cite{Sugiyama:2002tf,Hyun:2002wu,Hyun:2002wp,Gopakumar:2002dq,Das:2003yq}
(see~\cite{gb} for a different approach).
The Matrix String Theory
is characterized by a dimensionless combination $Mg_s$ of the strength
of the flux $\mu$, the string coupling $g_s$ and the radius $R$,
\begin{equation} \label{defM}
Mg_s = \frac{\mu l_s^2}{R} g_s.
\end{equation}
When $Mg_s \ll 1$ the excitations are standard Type IIA strings.
For $Mg_s \gg 1$ the dominant excitations include fuzzy spheres,
properties of which have been studied in detail in
\cite{Das:2003yq,Michelson:2004fh}. (Fuzzy spheres and other membranes 
in \ppwave\ Matrix Theory and Matrix String Theory have also been
studied in
\cite{Dasgupta:2002hx,Dasgupta:2002ru,Shin:2003np,Chen:2003sm,Furuuchi:2003sy,%
Janssen:2004jz,Lee:2004kv,Lozano:2005kf,jhp}.)

Note that $M g_s \gg 1$ is a weak coupling regime in the Yang-Mills
theory~\cite{Dasgupta:2002hx}.  This permits a substantial amount of
control over the theory, even though the corrresponding string theory
is not perturbative in this domain.  In particular, we can use this to
explore the region near the initial time singularity.

In this paper we show that \ppwave\ backgrounds can be easily modified
to yield \ppwave{s} with null-like linear dilatons, leading to a time
dependent coupling constant. Following steps similar to
\cite{Craps:2005wd} we then write down Matrix String Theories in
these backgrounds.
As time flows, the parameter which controls the Matrix String Theory
flows from weak to strong coupling. Consequently at early times
fuzzy spheres dominate the spectrum, while at late times the
spectrum consists of ordinary perturbative strings.

Earlier work on cosmology using Matrix Theory include,
\eg~\cite{Alvarez:1997fy,Freedman:2004xg}.
Following the work of \cite{Craps:2005wd}, 
other backgrounds with linear dilatons have
been studied in \cite{Li:2005sz,Li:2005ti}. 
Time-dependent \ppwave\ backgrounds have been studied
for a long time. Relevant papers include
\cite{prt,bopt}.
In particular,
Ref.~\cite{prt} discussed big bang-type singularities in the context
of {\em perturbative\/} strings in \ppwave\ backgrounds with a
nontrivial dilaton.

\section{Null Linear Dilaton \ppWave{s}}

In this section we give examples of \ppwave\ solutions of M theory which
lead to
string theory backgrounds with null linear dilatons.

\subsection{Maximally supersymmetric \ppwave}

Perhaps the simplest and most useful example of \ppwave\ solutions
with a linear dilaton arise from the maximally supersymmetric
M-theory \ppwave. The standard solution of \cite{Berenstein:2002jq}
can be easily modified to yield the eleven
dimensional background,
\bea \label{maxppdil}
ds^2 & = & e^{2 Q x^+/3} \{ 2 dx^+ dx^-  - H(x^+,x^a,x^{a''})~
(dx^+)^2
- \frac{2\mu(x^+)}{3} e^{-Q x^+} x^8 dx^9 dx^+
+ \sum_{i=1}^8 (dx^i)^2 \} \nn \\
& &  + ~e^{-4 Q x^+/3} ~(dx^9)^2, \nn \\
^{(4)}F & = & \mu(x^+) e^{Q x^+}~ dx^+ \wedge dx^1 \wedge dx^2 \wedge dx^3,
\eea
where 
\ben \label{defH}
H (x^+,x^a,x^{a''}) \equiv ( \frac{\mu(x^+)}{3} )^2 \sum_{a=1}^3 (x^a)^2
       + ( \frac{\mu(x^+)}{6} )^2 \sum_{a''=4}^7 (x^{a''})^2,
\een
and
we have used indices $a=1,2,3$ and
$a''=4,5,6,7$. Compactification along $x^9$ then yields a IIA solution
with a dilaton of the form (\ref{nulldilaton}).

For $Q=0$, and constant $\mu(x^+)$,
this is the maximally supersymmetric M-theory \ppwave, 
written in coordinates amenable to compactification along $x^9$
\cite{Sugiyama:2002tf,Hyun:2002wu,Das:2003yq}.
This remains a solution for $Q \neq 0$ and for
any function $\mu(x^+)$.
To see this, first note that the four-form
squares to zero and
is closed.  It is also co-closed
since the Hodge dual of $dx^+$ is aligned with $dx^+$.

To verify Einstein's equations it is convenient to choose the elfbein
(the caret denotes a tangent space index)
\begin{equation}
\begin{gathered}
e^{\hat{-}} = e^{Q x^+/3} \left\{ dx^- 
- \tfrac{1}{2} \left[ \left(\tfrac{\mu(x^+)}{3}\right)^2 (x^a)^2
                      + \left(\tfrac{\mu(x^+)}{6}\right)^2 (x^{a''})^2
                      + \left(\tfrac{\mu(x^+)}{3}\right)^2 (x^8)^2\right] dx^+
\right\},
\\
\begin{aligned}
e^{\hat{+}} &= e^{\frac{Q}{3} x^+} dx^+, &
e^{\hat{a}} &= e^{\frac{Q}{3} x^+} dx^a, &
e^{\hat{a}''} &= e^{\frac{Q}{3} x^+} dx^{a''}, &
e^{\hat{8}} &= e^{\frac{Q}{3} x^+} dx^8, &
\end{aligned} \\
e^{\hat{9}} = e^{-2 Q x^+/3} dx^9 - \frac{\mu(x^+)}{3} e^{Q x^+/3} x^8 dx^+.
\end{gathered}
\end{equation}
%
From this one finds that the nonvanishing components of the Riemann
tensor are,
\begin{equation}
\begin{gathered}
\begin{aligned}
R_{+a+a} &= \tfrac{\mu^2(x^+)+Q^2}{9} e^{2 Q x^+/3}, &
R_{+a''+a''} &= \tfrac{\frac{1}{4}\mu^2(x^+)+Q^2}{9} e^{2 Q x^+/3}, &
R_{+8+8} &= \tfrac{\frac{1}{4} \mu^2(x^+)+Q^2}{9} e^{2 Q x^+/3},
\end{aligned} \\ \begin{aligned}
R_{+9+9} &= \frac{\frac{1}{4}\mu^2(x^+)- 8 Q^2}{9} e^{-4 Q x^+/3}, &
R_{+8+9} &= \frac{2\mu(x^+) Q - \mu'(x^+)}{6} e^{- Q x^+/3},
\end{aligned}
\end{gathered}
\end{equation}
from which one quickly obtains the sole nonvanishing component of the
Ricci tensor
\begin{equation}
R_{++} = \frac{\mu^2(x^+)}{2}
 = \frac{1}{12} F_{+\mu\nu\rho} F_+{^{\mu\nu\rho}},
\end{equation}
so that indeed Einstein's equations are satisfied.

Another way to see that \eqref{maxppdil} is a solution 
is to consider the Type IIA solution
that corresponds to (\ref{maxppdil}). The string frame 
metric,
dilaton and Ramond-Ramond (RR) fields are given by
\begin{equation} \label{maxdil2a}
\begin{gathered}
ds^2 = 2dx^+dx^-
- [ H(x^+,x^a,x^{a''}) + (\frac{\mu (x^+)}{3})^2~(x^8)^2)] (dx^+)^2
+ \sum_{i=1}^8 (dx^i)^2, \\
\begin{aligned}
^{(4)}F_{+123} & = \mu (x^+)e^{Qx^+}, \qquad &
^{(2)}F_{+8} & = \frac{\mu (x^+)}{3}e^{Qx^+}, \qquad &
\Phi & = - Qx^+.
\end{aligned}
\end{gathered}
\end{equation}
It is well known that the IIA equations of motion are satisfied
for $Q=0$ and for any function $\mu (x^+)$. The equations of motion
involve the dilaton in two ways. First there are explicit terms
involving derivatives,%
\footnote{We use $\mu,\nu,\dots$ for 10-dimensional indices;
$A,B,\dots$ for 11-dimensions and $\alpha,\beta,\dots$ for $1+1$-dimensions.}
$\nabla_\mu \nabla_\nu \Phi, (\nabla \Phi)^2,
\nabla^2 \Phi$. Since the metric has $g^{++} = 0$,
and the Christoffel symbols have $\Gamma^+_{\mu\nu} = 0$, it is trivial
to see that all these derivative terms vanish. The second class of terms
where the dilaton
appears are as overall coefficients $e^{2\Phi}$ multiplying the RR
field strength terms in the Einstein equations,
namely $e^{2\Phi} (^{(4)}F)^2$
and $e^{2\Phi} (^{(2)}F)^2$. However, the factors
of $e^{Qx^+}$ in the field strengths in (\ref{maxdil2a}) precisely
cancel these coefficients. Therefore the Einstein and dilaton equations
reduce precisely
to those at $Q=0$.
Finally 
the field strengths are co-closed
since they depend only on $x^+$ and all
 components $F^{+\mu\nu\alpha}$ or $F^{+\mu}$ vanish. 
We have compiled the Killing vectors of the IIA background in
Appendix~\ref{sec:killing}.

\subsubsection{Supersymmetry} \label{sec:maxsusy}

It is by now well known~\cite{fp,ghpp,clp,clp2}
that all \ppwave{s} are invariant under the sixteen so-called\cite{clp}
standard supersymmetries that satisfy $\Gamma^{\hat{+}} \epsilon = 0$.
These supersymmetries are also preserved by the null linear dilaton.
Explicitly,
the Killing spinor equation is
\begin{equation} \label{killspin}
\cD_A \epsilon = \nabla_A \epsilon - \Omega_A \epsilon = 0,
\end{equation}
where, of course, $\nabla_A$ is the usual covariant derivative, and
\begin{equation} \label{defOmega}
\Omega_A = \frac{1}{24 \cdot 4!} {^{(4)}F}_{BCDE} 
 \left(3 \Gamma^{BCDE} \Gamma_A - \Gamma_A \Gamma^{BCDE}\right).
\end{equation}
%
For the solution~\eqref{maxppdil}, the sixteen standard supersymmetries
are
\begin{equation} \label{maxppss}
\epsilon(x^+)
= \exp \left[\frac{Q}{6} x^+ - \left(\int^{x^+} dx \mu(x)\right)
      \left( \tfrac{1}{4} \Gamma^{123} + \tfrac{1}{12} \Gamma^{89} \right)
 \right]
 \epsilon_0,
\qquad \Gamma^+ \epsilon_0 = 0.
\end{equation}

However, it must be emphasized that the standard
supersymmetries~\eqref{maxppss} are not interesting.  Their existence
does not guarantee the supergravity equations of motion, and they are
only nonlinearly realized in the Matrix Theory, and on the
Green-Schwarz worldsheet.

The interesting supersymmetries are the supernumerary
supersymmetries---that is, those supersymmetries that are not
standard.  For example, the existence of a single
supernumerary supersymmetry {\em does\/} guarantee the supergravity equations
of motion (a short proof of this can be found in~\cite{Michelson:2004fh}).
Also, the supernumerary supersymmetries are linearly
realized in the Matrix Theory and on the worldsheet, and are therefore
useful for analyzing the spectrum and for nonrenormalization theorems.
It is also interesting that toroidal compactification usually breaks
some, if not all, of the supernumerary supersymmetry.\cite{Michelson:2002wa}
For example,
compactifying the maximally supersymmetric \ppwave\ to the IIA theory
breaks
eight of the sixteen supernumerary supersymmetries.%
\cite{Sugiyama:2002tf,Hyun:2002wu,Das:2003yq}

Unfortunately, it is also well known 
that the existence of supernumerary supersymmetries for a
\ppwave\ requires that the coefficients in the (Brinkmann) metric be
constant.  Therefore, there are no supernumerary supersymmetries for
the solution~\eqref{maxppdil}.  (Indeed, this is straightforward to
verify by considering, say,  the integrability condition 
$\com{\cD_+}{\cD_a} = 0$.)
Nevertheless, it will turn out that, for a particular choice of $\mu(x^+)$,
the Matrix String Theory will have linearly realized supersymmetry currents,
which we expect will be useful.

\subsection{Linear dilaton dressing of a 24 supercharge \ppwave} \label{sec:24}

Similar solutions can be written down which have a smaller number of
supersymmetries in the $Q=0$ limit.  For example, one can consider the
Penrose limit of the NS-NS flux-supported
IIA AdS$_3 \times $S$^3$ background.
The sixteen
supersymmetries of the AdS background are enhanced to 24 upon taking
the Penrose limit.
This geometry was previously considered
in~\cite{clp,clp2,rt,gms,Michelson:2004fh}.
Upon adding the linear dilaton,
one obtains the M-theory metric and the four form,
\begin{equation} \label{AdS3dil}
\begin{aligned}
ds^2 &= e^{2 Q x^+/3} \left[2 dx^+ dx^- - \mu^2 \sum_{a=1}^4 (x^a)^2 (dx^+)^2
 + \sum_{I=1}^8 (dx^I)^2 \right] + e^{-4 Q x^+/3} (dx^9)^2, \\
^{(4)}F &= 2 \mu dx^+ \wedge \left[ dx^1 \wedge dx^2 + 
 dx^3 \wedge dx^4 \right] \wedge dx^9.
\end{aligned}
\end{equation}
It may be checked that the equations of motion
are satisfied.  Note that for this \ppwave, the linear dilaton does not appear
in the field strength; this is related to the fact that the corresponding 
IIA configuration is supported by NS-NS flux, and not RR flux(es).

\subsubsection{Supersymmetry} \label{sec:24susy}

Identical comments apply here as
in~$\S$\ref{sec:maxsusy}.  There are sixteen standard supersymmetries,
$\epsilon(x^+)
= \exp \left[\frac{Q}{6} x^+ - \int^{x^+} dx \frac{\mu(x)}{2}
 (\Gamma^{129} + \Gamma^{349})
 \right]
 \epsilon_0,
\Gamma^+ \epsilon_0 = 0$,
and no supernumerary supersymmetries for the
configuration~\eqref{AdS3dil}, unless $Q=0$ and $\mu$ is constant.

\subsection{Linear dilaton dressing of the 26 supercharge \ppwave}
\label{sec:26}

The nonperturbative physics~\cite{Michelson:2004fh} of the 26
supercharge \ppwave~\cite{jm26} is similar to that of the maximally
supersymmetric \ppwave.  However, the corresponding IIA configuration,
and therefore the Matrix String Theory, of the 26 supercharge \ppwave\
has two more supercharges than does the (compactification of the)
maximally supersymmetric \ppwave.  One can dress the 26 supercharge
\ppwave\ with a dilaton in a way that precisely
mimics~\eqref{maxppdil}.  The resulting configuration is,
\begin{equation}
\begin{aligned}
ds^2 &= e^{2 Q x^+/3} \Bigl\{ 2 dx^+ dx^- 
   - \mu^2(x^+) \sum_{i=1}^7 (x^i)^2 (dx^+)^2
   + \sum_{i=1}^7 (dx^i)^2 + (dx^8)^2
\\ & \qquad
+ 2 \mu(x^+) x^8 e^{-Q x^+} dx^+ dx^9 \Bigr\}
+ e^{-4 Q x^+/3} (dx^9)^2,
\\
{^{(4)}}F &= \mu(x^+) e^{Q x^+} dx^+ \wedge
\left[ -3 dx^1 \wedge dx^2 \wedge dx^3
   - \sum_{y=1}^3 dx^y \wedge \omega_y^- \right],
\end{aligned}
\end{equation}
where $\omega_y^-$ are the anti-selfdual two-forms of the $\ZR^4$
spanned by $x^{4\cdots7}$.

\subsection{The nature of the singularity}

For the configurations introduced in this section, the dilaton blows
up at $x^+=-\infty$.  Since $g^{++}=0$, the surface $x^+ = -\infty$
is null-like.
The geodesic equations include
\begin{equation}
\frac{d^2 x^+(\lambda)}{d\lambda^2}
 = - \frac{2 Q}{3} \left(\frac{d x^+(\lambda)}{d\lambda}\right)^2,
\end{equation}
where $\lambda$ is the affine parameter, so that either
$x^+=\text{constant}$ or
\begin{equation}
x^+(\lambda) = \frac{3}{2Q} \ln \lambda,
\end{equation}
(by choice of affine parameterization).  This demonstrates that the
null singularity at \hbox{$x^+=-\infty$} is reached in finite affine
parameter, (and that reaching $x^+ = \infty$ requires infinite affine
parameter, and is therefore not a singularity.)  These conclusions
are precisely as
in~\cite{Craps:2005wd}---and, in fact, are independent of $\mu$!

Let us also point out that all time-like geodesics start at the singularity.
For example, for the \ppwave~\eqref{maxppdil},
the only geodesics that do not hit the singularity have the very simple form
\begin{align}
x^+ &= x_0^+, &
x^i &= x_0^i + p^i \lambda, \quad i \neq 9, &
x^9 &= x_0^9, &
x^- &= x_0^- + p^- \lambda + \frac{Q}{6} (p^i)^2 \lambda^2,
\end{align}
which is never timelike, and null iff $p^i = 0$---and still
independent of $\mu$.

\section{Matrix String Theory}

Following \cite{Craps:2005wd} we propose that the
action of Matrix String Theory in the null linear dilaton - \ppwave\ 
backgrounds may be obtained from the usual action by replacing the
string coupling $g_s$ by $\bg_s$ defined in (\ref{bargs}). 
The resulting action, for the \ppwave~\eqref{maxppdil}, is
($i,j=1\dots 8$)
\bea
S = \int d\tau d\sigma \Tr & \{ &
\frac{1}{2} \bg_s^2 F_{\tau \sigma}^2
+ \frac{1}{2} (D_\tau X^i)^2 
- \frac{1}{2} (D_\sigma X^i)^2
+ \frac{1}{4 \bg_s^2} \com{X^i}{X^j}^2 \nn \\*
& & - \frac{1}{2} \left(\frac{M(\tau)}{3}\right)^2 (X^a)^2
-\frac{1}{2} \left(\frac{M (\tau)}{6}\right)^2 (X^{a''})^2 \nn \\*
& & - \frac{M (\tau)}{3} \bg_s X^8 F_{\tau\sigma}
- i \frac{M (\tau)}{3 \bg_s} \epsilon_{abc} X^a X^b X^c
\} + S_{\text{fermion}}.
\label{ppmatrixstring}
\eea
where
$M (\tau) $ is given in terms of $\mu (\tau)$ by 
\begin{equation}
M(\tau) = \frac{\mu (\tau) l_s^2}{R}.
\end{equation}
The worldsheet time $\tau$ has to be identified with $x^+$.

The fermionic part of the action is
\begin{equation}
S_{\text{fermion}} = \int d\tau d\sigma\, \Tr \left\{ 
-i\bar{\Psi} \fs{D} \Psi
+ \frac{1}{\bg_s} \bar{\Psi} \rho^3 \Gamma^i \com{X^i}{\Psi}
-i\frac{M}{4} \bar{\Psi} \rho^3 (\Gamma^{123} - \frac{1}{3} \Gamma^{89}) \Psi
\right\},
\end{equation}
where $\Psi$ is a $1+1$ dimensional Majorana fermion, each Weyl component of
which is an $U(N)$ matrix-valued Majorana-Weyl SO(8) spinor.  The SO(8)
chirality is opposite to the SO(1,1) chirality.%
\footnote{We use $\rho^3=-\rho^0 \rho^1$.}
The $\Gamma$'s
are SO(8) Dirac matrices, the $\rho$'s are SO(1,1) Dirac matrices,
and \hbox{$\fs{D} \equiv \rho^\tau D_\tau + \rho^\sigma D_\sigma$}.

We emphasize that we have not {\em derived} the above matrix
model using some kind of Seiberg-Sen~\cite{seibergsen,senseiberg} argument.
(Indeed, the accepted
derivation~\cite{Dasgupta:2002hx} of the $Q=0$ Matrix Model
is still suspect because of the use of a non-spacelike circle, whereas the
whole point of the Seiberg-Sen argument is to rephrase the null-like
compactification in terms of a {\em spacelike\/} one.  But it is not obvious
how to fix that derivation.) 

The bosonic terms in the above action may be, however, obtained by
considering the light-cone gauge action of a single massless particle
in 11 dimensions with momentum $p_-=N/R$ (or equivalently the action
for $N$ D0 branes in the IIA theory) following
\cite{Myers:1999ps,TV,preTV}. The lagrangian of the matrix theory may be
written as
\bea
L  =  \Tr & \{ & \frac{1}{2} (R G^{+-})G_{IJ}D_\tau X^I D_\tau X^J
 - \frac{1}{R}
G_{IJ}G^{-I} D_\tau X^J 
+\frac{1}{2G^{+-}R}( G_{IJ}G^{-I}G^{-J}-G^{--} ) \nn \\
& + & \frac{R}{4G^{+-}}G_{IK}G_{JL}[X^I,X^J][X^K,X^L]
- \frac{i}{2} A_{+IJ}X^I X^J \},
\label{generalaction}
\eea
where $G$ denotes components of the eleven dimensional metric and
$I,J= 1\cdots 9$ denote all the transverse directions. In writing
(\ref{generalaction}) we have used the fact that $G^{++}=0$ and that
the only nonvanishing components of the 3-form potential are of the
form $A_{+IJ}$. Using the metric and the potential from (\ref{maxppdil})
this becomes
\bea
L  =  R \Tr
& \{ & \frac{1}{2R^2}(D_\tau X^i)^2 + \frac{1}{2R^2}~e^{-2Q\tau} 
(D_\tau X^i)^2 - \frac{\mu}{3R^2}~e^{-Q\tau}~X^8 D_\tau X^9 \nn \\
&  - & 
\frac{1}{2}(\frac{\mu}{3R})^2(X^a)^2 -
\frac{1}{2}(\frac{\mu}{6R})^2(X^{a"})^2 \nn \\
& + & \frac{1}{4}e^{2Q\tau} [X^i,X^j]^2 +\frac{1}{2} [X^9,X^i]^2 - 
i\frac{\mu}{3R}
\epsilon_{abc}e^{Q \tau} X^aX^bX^c \}.
\label{mtheoryaction}
\eea
Simultaneous compactification and T-dualization along $X^9$ means we
introduce the spatial direction $\sigma$ and identify
\ben
X^9 = i{\hat R}D_\sigma, \qquad \Tr \rightarrow \int_0^{2\pi} d\sigma,
\qquad D_\tau X^9 = -{\hat R} F_{\tau\sigma}.
\een
A standard set of rescalings then leads to (\ref{ppmatrixstring}).
These steps are identical to those in \cite{Das:2003yq} and will not
be repeated here.  In a later section we will see that a comparison of
the dynamics of giant gravitons and those of fuzzy spheres provides an
important check on our guess.

To understand the nature of the big bang singularity in this model it is
useful to rewrite the action as if it had the background
metric~\eqref{conmetric}.  This effectively removes the
$\tau$-dependence from the string coupling.  In this form the action
is
\begin{multline}
S =  S_{\text{fermion}} + \int d\tau d\sigma\, \sqrt{-g} \Tr \left\{
- \frac{1}{4} g_s^2 F_{\alpha\beta}^2
- \frac{1}{2} (D_\alpha X^i)^2 
+ \frac{1}{4 g_s^2} \com{X^i}{X^j}^2
\right. \\ \left.
- \frac{1}{2} \left(\frac{\tM (\tau) }{3}\right)^2 (X^a)^2
-\frac{1}{2} \left(\frac{\tM (\tau) }{6}\right)^2 (X^{a''})^2
- \frac{\tM (\tau) }{6} g_s X^8 \epsilon^{\alpha\beta} F_{\alpha\beta}
- i \frac{\tM (\tau)}{3 g_s} \epsilon_{abc} X^a X^b X^c
\right\},
\end{multline}
where we have defined
\begin{equation}
\tM(\tau) \equiv M(\tau) e^{-Q\tau}.
\label{mass}
\end{equation}
Upon rescaling the fermions $\Psi_\text{new} = e^{-Q \tau/2}
\Psi_\text{old}$ (as for a conformal transformation) one has
%
\begin{equation}
S_{\text{fermion}} = \int d\tau d\sigma\,\sqrt{-g} \Tr \left\{ 
i\bar{\Psi} \fs{D} \Psi
 +
\frac{1}{g_s} \bar{\Psi} \rho^3 \Gamma^i \com{X^i}{\Psi}
-i\frac{\tM(\tau)}{4} \bar{\Psi} \rho^3
      (\Gamma^{123} - \frac{1}{3} \Gamma^{89}) \Psi
\right\}.
\end{equation}
In this form, we have a $1+1$ dimensional Yang-Mills
theory with a time-dependent mass for the scalars and 
constant coupling, defined on a space-time with metric
$g_{\alpha\beta}$ given in \eqref{conmetric}.

\subsection{Supersymmetry}

The Matrix String
action has nonlinearly realized supersymmetry
corresponding to the standard supersymmetries.  This is not interesting.

Generically, the action has no linearly realized supersymmetries (see
Appendix~\ref{sec:genSUSY}).  This is not surprising, since the
supergravity background has no supernumerary supersymmetries. The
supersymmetry variation of the action has two terms. The first, which
we call the bulk term, is an integral over $\sigma,\tau$, while the
second is a boundary term. At finite times, both these terms are
non-vanishing---the boundary term is non-vanishing, as in~\cite{Craps:2005wd},
because of
non-trivial boundary conditions on fermions in the $\sigma$
direction. The integrand of the bulk term however, vanishes
exponentially fast at late times while at the same time the radius of
the $\sigma$ circle also increases exponentially fast. This means that
in some sense supersymmetry is restored at late times.

Interestingly,
for the special case,
\begin{equation} \label{special}
\mu(\tau) = \mu_0 e^{Q \tau},
\end{equation}
the Matrix String Theory has 8 linearly realized supercurrents
corresponding to the transformations
\begin{subequations} \label{MppSUSY}
\begin{gather} \label{MppSUSYX}
\begin{align} 
\delta X^i &= i \bar{\Psi} \rho^3 \Gamma^i \epsilon, &
\delta A_\alpha &= i \bar{\Psi} \rho_\alpha \epsilon, &
\end{align} \\ \label{MppSUSYPsi}
\begin{aligned}
\delta \Psi &= \frac{1}{4} F_{\alpha\beta} \rho^{\alpha\beta} \epsilon
 - \frac{1}{2} \epsilon^\alpha{_\beta} 
   D_\alpha X^i \Gamma^i \rho^\beta \epsilon
 + \frac{i}{4} \com{X^i}{X^j} \Gamma^{ij} \epsilon
\\ & \qquad
 + \frac{\tM(\tau)}{6} X^a \Gamma^a \Gamma^{123} \epsilon
 - \frac{\tM(\tau)}{12} X^{a''} \Gamma^{a''} \Gamma^{123} \epsilon,
\end{aligned}
\intertext{where} \label{epsMppSUSY}
\epsilon = \Gamma^{4567} \epsilon_0 = \Gamma^{12389} \epsilon_0.
\end{gather}
\end{subequations}
Note that for $Q\neq 0$, these supercurrents cannot be integrated,
because of the nonvanishing boundary term.  (The boundary terms are
given in Eq.~\eqref{SUSY:td}.)  Of course, for $Q=0$ these do yield
supersymmetries that are precisely those given and discussed in
\cite{Das:2003yq}. For the special case~\eqref{special} however, the
string frame metric of the Type IIA theory evolves from {\em flat}
space at early times to a space-time whose geodesics at late times are
effectively those of a two-dimensional space-time, since the $++$
component of the metric provides a harmonic restoring force towards
$x^i=0$ which becomes infinite as $x^+ \rightarrow \infty$.%
\footnote{This discussion, however, is for the sector of the theory
with constant \hbox{$P_- = N/R$}.  The coordinate change \hbox{$y^+ =
e^{Q x^+}/Q, y^- = e^{-Q x^+} x^-$}, results in the metric
\hbox{$ds^2 = 2 dy^+ dy^- - [(\frac{\mu_0}{3})^2 (x^a)^2 
+ (\frac{\mu_0}{6})^2 (x^{a''})^2 + (\frac{\mu_0}{3})^2 (x^8)^2 + 2
 \frac{y^-}{y^+} ] (dy^+)^2 + (dx^i)^2$}, which asymptotes at large $y^+$
to the usual \ppwave.}

\section{Fuzzy Spheres}

Let us recall the ground states of the \ppwave\ Matrix string theory
for $Q=0$ and constant $\mu (\tau) = \mu$. As shown in
\cite{Das:2003yq}, one can rescale the coordinates and the fields by
\begin{equation}
\tau \rightarrow M\tau, \qquad \sigma \rightarrow M\sigma, \qquad
A_\mu \rightarrow A_\mu / M, \qquad X^i \rightarrow X^i/(Mg_s),
\qquad \Psi \rightarrow \Psi/(M^{3/2}g_s),
\label{rescaling}
\end{equation}
to bring out an overall factor of $(Mg_s)^2$ out of the action, with
no other factor of $M$ or $g_s$ inside. This shows that the quantity
$Mg_s$ is the inverse coupling constant of the theory. 
For $Mg_s \ll
1$ the theory is strongly coupled, and only fields which lie in 
the Cartan subalgebra survive. By a gauge choice we can eliminate the
gauge field and choose the $X^i$ to be diagonal; 
the remaining gauge freedom allows boundary conditions as we go around
the $\sigma$ direction that correspond to multi-string states.
This leads to usual perturbative IIA theory.

For $Mg_s \gg 1$, the dominant configurations are obtained by solving
the classical equations of motion and the lowest energy states include
static fuzzy spheres. In terms of the original variables, these
solutions are given by
\begin{equation}
X^a = \frac{Mg_s}{3}J^a, \qquad (a=1,2,3),
\end{equation}
where $J^a$ are generators of a $N$-dimensional representation of
$SU(2)$ algebra. Note that these are independent of $\sigma$; they 
should be called ``fuzzy cylinders'' in the YM theory. 
As discussed in \cite{Das:2003yq}, small fluctuations around this ground 
state have an equally spaced spectrum of $N^2$ degrees of freedom whose
boundary conditions along the $\sigma$ direction correspond
to fuzzy cylinders that do not generically close upon going around
the $\sigma$ circle.

The rescaling in (\ref{rescaling}) can be performed in the matrix
model in (\ref{ppmatrixstring}) as well (note the quantity that
appears in the rescaling is $g_s$ and not $\bg_s$). However now
various terms in the action have explicit time dependence. Thus $Mg_s
\gg 1$ does not guarantee that classical solutions dominate for all
times.  In fact it is clear from (\ref{ppmatrixstring}) that as $\tau
\rightarrow \infty$ the coefficients of the terms which involve
commutators of $X$ grow large, so that the fields $X^i$ are more and
more constrained to lie in the Cartan subalgebra even when $Mg_s \gg
1$. It may be easily checked that in this case the matrix string
action reduces to the standard Green-Schwarz light cone worldsheet
action of a string in this background.  Our aim to is investigate the
dynamics of these fuzzy spheres as the geometry evolves from the big
bang.

\section{Giant Gravitons}

In the $N = \infty$ limit of the $Q = 0$ background, 
the fuzzy cylinders of the Matrix String Theory become
giant gravitons \cite{McGreevy:2000cw,Grisaru:2000zn,Hashimoto:2000zp,%
Das:2000fu,Das:2000ab,Das:2000st}. 
In the IIA description these are $D2$ branes with quantized
momenta $p_- =N/R$ along $x^-$, while in the M-theory lift they are $M2$ 
branes with this momentum. To get an idea of the nature of excitations
of string theory for $Q \neq 0$ it is useful to re-examine the
dynamics of these $M2$ branes in the background given by
(\ref{maxppdil}). We will do the analysis for a general $\mu (X^+)$.

We will use polar coordinates $(r,\theta,\phi)$ in the plane defined by
the Cartesian coordinates $x^a, a = 1\cdots 3$. In these coordinates
the field strength is 
\begin{equation}
F_{+r\theta\phi} = \mu(X^+) e^{Q X^+} r^2 \sin \theta,
\end{equation}
and we can choose a gauge such that the potential is
\begin{equation}
C_{+\theta\phi} = \frac{1}{3} \mu(X^+) e^{Q X^+} r^3 \sin\theta,
\end{equation}
with all other components zero. 

The M2 brane is wrapped around the $(\theta,\phi)$ direction.
This allows us to fix a gauge in which the spatial coordinates on the
worldvolume are identified with the angles $\theta$ and $\phi$. We
want to restrict our dynamics to the sector where the remaining
coordinates are independent of $\theta,\phi$. This is a consistent
truncation which respects the equations of motion. Then all
worldvolume fields depend only on the worldvolume time $\tau$ and the
action of the M2 brane reduces to that of a point particle:
\begin{equation}
S = 4\pi T_2\int d\tau\left[- e^{2 Q X^+/3}r^2(\tau){\sqrt{-G_{AB}\partial_\tau
X^A \partial_\tau X^B}} + \frac{\mu(X^+)}{3} e^{Q X^+} r^3(\tau) \partial_\tau
X^+\right], 
\label{eq:bfour}
\end{equation}
where $A,B$ stand for $(r,X^{a''},X^8, X^9, X^\pm)$. The metric
$G_{AB}$ is the solution given in (\ref{maxppdil}).

The canonical momenta are given by
\begin{equation}
P_A = 4\pi T_2 \left[ \frac{e^{2 Q X^+/3}}
{{\sqrt{-G_{AB}\partial_\tau
X^A \partial_\tau X^B}}} r^2 G_{AB}\partial_\tau X^B +
\frac{\mu(X^+) r^3}{3} e^{Q X^+} \delta_A^+ \right].
\label{eq:bsix}
\end{equation} 
The identity 
\begin{equation}
G^{AB}\left[P_A - \frac{4\pi T_2}{3}\mu(X^+) e^{Q X^+} r^3 \delta_A^+\right]
\left[P_B - \frac{4\pi T_2}{3}\mu(X^+) e^{Q X^+} r^3 \delta_B^+\right]
= -(4\pi T_2 r^2)^2 e^{4 Q X^+/3},
\label{eq:bseven}
\end{equation}
yields the dispersion relation.

The above gauge choice still allows arbitrary reparametrization of the
worldvolume time coordinate $\tau$. We  fix this by choosing a gauge
$\tau = X^+$ so that the canonical Hamiltonian is given by 
$ H = - P_+$.

Since $X^-$ and $X^9$ are isometry directions, $P_-$ and $P_9$ are 
conserved quantities. We will consider the sector of the theory in which
$ P_9 = 0$ .
The momenta $P_{a''}$ and $P_8$ are not conserved.
However it is easy to see that 
$ P_8=X^8=P_{a''}=0$
is a solution of the equations of motion. Furthermore, in this sector
$ X^{a''}=X^9=0 $
is a consistent solution. The Hamiltonian of the truncated problem then
simplifies considerably and may be written as
\begin{equation}
H = \frac{P_r^2}{2P_-}
+\frac{(4\pi T_2)^2}{2P_-}~e^{2Q\tau}~\left(\tfrac{\mu(\tau)P_-}{12\pi
T_2}~r~e^{-Q\tau}
 - r^2 \right)^2,
\label{hamilt}
\end{equation}
where we have used the explicit form of the metric
(\ref{maxppdil}). This is the Hamiltonian of a particle with mass
$P_-$ moving in a time-dependent potential.
The equation of motion is
\begin{equation}
P_-^2 \frac{d^2 r}{d\tau^2} + (4\pi T_2)^2~e^{2Q\tau}~
\left(\tfrac{\mu(\tau)P_-}{12\pi
T_2}~r~e^{-Q\tau}
 - r^2 \right)
\left(\tfrac{\mu(\tau)P_-}{12\pi
T_2}~e^{-Q\tau}
 - 2r  \right)^2.
\label{eqmotion}
\end{equation}
For $Q=0$ and constant $\mu$, the lowest energy solution is static with 
\begin{equation} \label{prevgiant}
r = r_0 = \frac{\mu}{12\pi T_2}P_-.
\end{equation}
This is the giant graviton solution discussed in \cite{Das:2003yq}.

For $Q \neq 0$ it is useful to introduce the variables
\begin{equation}
\Rho (t) \equiv \frac{12\pi T_2}{\mu(\tau)
  P_-}r(\tau) e^{Q \tau}, \qquad t = Q \tau,
\end{equation}
The equation \eqref{eqmotion} then becomes
\begin{equation} \label{Rhoeq}
\frac{d^2 \Rho}{dt^2}
- 2 \frac{d \Rho}{dt} \left(1 - \frac{1}{2} \frac{A'(t)}{A(t)} \right)
+ \Rho \left(1 -\frac{A'(t)}{A(t)} - \frac{A'(t)^2}{4 A(t)^2}
             + \frac{A''(t)}{2 A(t)}
       \right)
+ A(t) (1-2\Rho)(\Rho-\Rho^2)
= 0,
\end{equation}
where
\begin{equation}
 A(t) \equiv (\frac{\mu(\tau)}{3 Q})^2,
\end{equation}
is the single parameter which enters into this equation.

\begin{figure}[tb]
\begin{center}
\includegraphics[width=3.0in,height=2.0in]{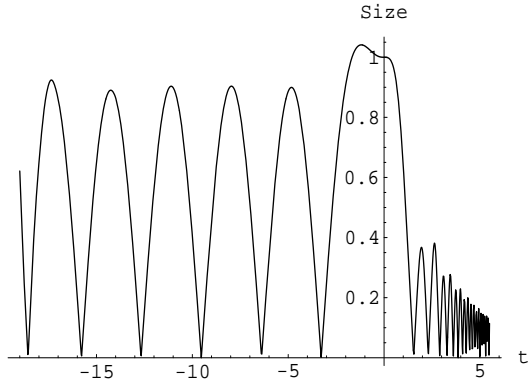}
\caption{The scaled size of a giant graviton
as a function of the time $t$ for $A=1$. 
\label{size}
}
\end{center}
\end{figure}

Let us first examine the nature of the solutions to (\ref{Rhoeq}) for constant
$\mu$, \ie\ a constant $A$.
In this case the differential equation~\eqref{Rhoeq} reads
\begin{equation} \label{RhoeqA}
\frac{d^2\Rho}{dt^2} - 2 \frac{d \Rho}{dt} 
+ \Rho\left[ (1+A) - 3 A \Rho + 2 A \Rho^2 \right] = 0.
\end{equation}
As this equation has no explicit $t$-dependence, it is easy to see that
for some range of the values of $A$ there is a solution with
constant $\Rho = \Rho_0$
\begin{equation}
\Rho_0 = 0, \frac{3}{4} \pm \sqrt{\frac{1}{16} - \frac{1}{2A}}.
\end{equation}
which are all real for $A > 8$.
This means that the radius of the giant graviton $r(\tau)$ has the
behavior
\begin{equation}
r(\tau) = \frac{\mu P_-\Rho_0}{12\pi T_2} e^{-Q\tau}.
\label{specialsol}
\end{equation}
In other words, at early times the giant gravitons are big, while
at late times they shrink to zero size at an exponential rate.
We see that for large $A$, this roughly reproduces (upon choosing the ``+''
root)~\eqref{prevgiant}, in the sense that one obtains
\begin{equation}
r(t) = \frac{\mu P_-}{12 \pi T_2} \left[ \frac{3}{4} 
          + \frac{1}{4} \sqrt{1-8 A^{-1}} \right] e^{-Q \tau}
= \frac{\mu P_-}{12 \pi T_2} e^{-Q \tau} + \order{A^{-1}}.
\end{equation}
For $A < 8$ the roots become complex and there are no nonzero constant $\Rho$
solutions. 

\begin{figure}[bt]
\begin{center}
\includegraphics[width=3.0in,height=2.0in]{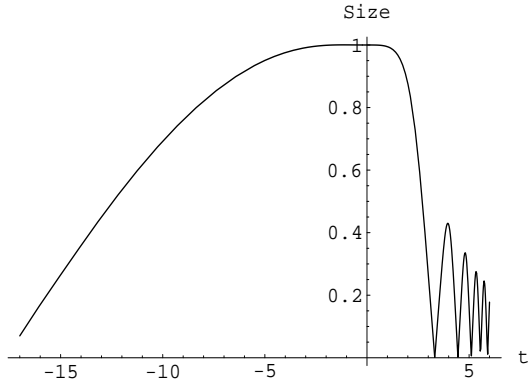}
\caption{Same as Fig.~\ref{size}, but with $A=0.01$.
Extended to smaller times, the size appears to hit a zero and then
rise rapidly as time is further decreased; however, this is a regime
for which our numerical solution is unreliable.
\label{size2}
}
\end{center}
\end{figure}

Note that (\ref{specialsol}) is an {\em exact\/} solution and not
an adiabatic approximation. However this corresponds to special
initial conditions.
Generically,
it is not hard to see that if one starts out with a M2 brane of finite
size at early times, it will collapse to zero size at late times
exponentially, regardless of the value of $A$.%
\footnote{As a crude approximation, one can use a ``perturbative expansion''
and ignore the higher powers of $\Rho$ in Eq.~\eqref{RhoeqA}.  Then
the solutions are $\Rho(t) \sim \exp[(-1 \pm i \sqrt{A}) t]$,
demonstrating that the fuzzy spheres are falling off even faster than
for large $A$, and oscillate with a period of order $A^{-1/2}$.}
Fig.~\ref{size} shows a numerical solution of this equation for
$A=1$, plotted in terms of the scaled size of the giant graviton
\begin{equation}
\rho (\tau) = \Rho (\tau) e^{-Q\tau},
\end{equation}
with initial conditions
$\rho(0)=1$ and ${\dot \rho}(0)=0$, (\ie\ a typical size at a
time when the coupling is $\order{g_s}$).
$\rho(t)$ oscillates in $t$, but the amplitude goes
down at large $t$.  This means that at late times the size goes to
zero essentially because of the fact that in equation (\ref{eqmotion})
the potential becomes infinitely steep. This behavior is typical
for other initial conditions---the behavior at early times depends on 
the initial conditions but the amplitude always decreases to zero at
large times. 
For smaller 
values of $A$ the size of the M2 brane remains big for a longer time
period, but the late time behavior is universal. This is illustrated
in Fig.~\ref{size2}.

\begin{figure}[tb]
\begin{center}
\includegraphics[width=3.0in,height=2.0in]{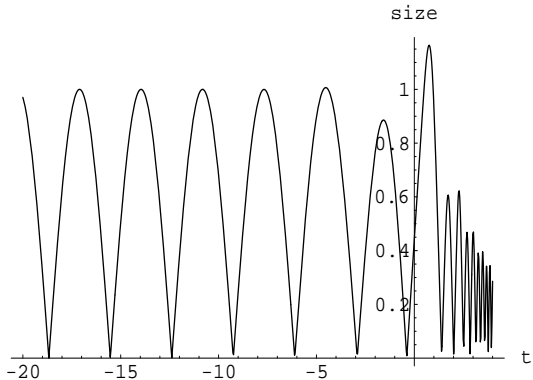}
\caption{Similar to Fig.~\ref{size}, with $A=1$, but initial conditions
at an earlier time.
\label{earlysize}
}
\end{center}
\end{figure}

\begin{figure}[bt]
\begin{center}
\includegraphics[width=3.0in,height=2.0in]{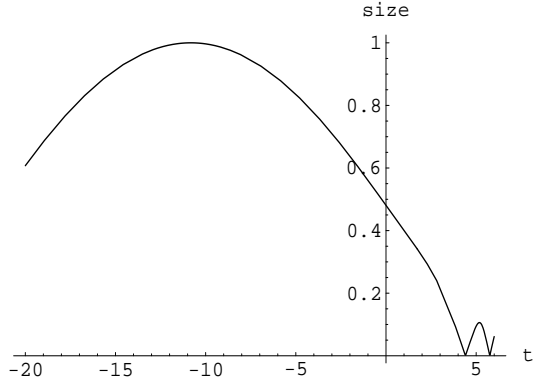}
\caption{Same as Fig.~\ref{earlysize}, but with $A=0.01$.
\label{earlysize2}
}
\end{center}
\end{figure}

Alternatively one could impose final state conditions demanding that
the size of the giant graviton vanishes at sufficiently late time. As
expected, typically the amplitudes of oscillations of the size
increases as we go back in time.  One can also demand that the giant
graviton have a typical size at an earlier time.  This is depicted in
Figs.~\ref{earlysize} and~\ref{earlysize2}.  One continues to see that the
size of the giant graviton falls off at late times.

Let us now discuss some features of the solution for a time-dependent
$\mu (\tau)$. We have not investigated this in full generality.
However it is easy to see that
when $\mu(\tau) = \mu e^{Q \tau}$, then one quickly finds static solutions
\begin{equation}
r(\tau) = r_0 = 0, \frac{\mu P_-}{12 \pi T_2}, \frac{\mu P_-}{24 \pi T_2}.
\end{equation}
In fact the first and second  solutions have vanishing energies at all times, just as
for $Q=0$ with constant $\mu$. 
For generic initial conditions, a numerical solution shows that the
size is attracted to either $r = 0$ or towards $r = r_0$ at late times.

\section{Dynamics of Fuzzy Spheres}

The dynamics of fuzzy sphere configurations is similar
to that of giant gravitons. To examine the simplest 
situation consider the matrix string action in the
sector
\begin{equation}
X^{4} = X^{5} = \cdots = X^8 = 0, \qquad A_\mu = 0,
\qquad \Psi = 0.
\end{equation}
This restriction is consistent with the equations of
motion. The action then takes a simple form
\begin{equation}
S = \int d\tau d\sigma \Tr \left\{ \frac{1}{2} (\partial_\tau X^a)^2 -
\frac{1}{2} (\partial_\sigma X^a)^2 + \frac{1}{4\bg_s^2}\left(
[X^a,X^b] - i\frac{M(\tau)\bg_s}{3} \epsilon^{ab}{_c} X^c \right)^2 \right\}.
\label{faction}
\end{equation}

We will look at $\sigma$-independent configurations of the form
\begin{equation}
X^{a} (\sigma,\tau)  =  S(\tau) J^a, \qquad (a=1,2,3), 
\end{equation}
where $J^a$ is an $N$-dimensional representation of $SU(2)$. When $Q=0$
this configuration would be a classical solution with a constant $S =
{Mg_s/3}$---this is the fuzzy cylinder of the IIB theory carrying a D1
brane charge, which becomes a fuzzy sphere of the original IIA theory
carrying a momentum $p_-$ along $x^-$. When $J^a$ is the irreducible
representation we have a single fuzzy sphere whereas a $J^a$ in a
reducible representation corresponds to multiple fuzzy spheres.

When $Q \neq 0$ a constant $S$ would not be a solution, and we would
like to find how $S$ evolves in time.  For $J^a$ in the $N$-dimensional
irreducible representation,
\begin{equation}
\Tr \sum_a (J^a)^2 = \frac{N (N^2-1)}{4},
\end{equation}
and the action (\ref{faction}) becomes
\begin{equation}
S = \frac{N (N^2-1)}{8}\int d\tau d\sigma [ (\frac{dS}{d\tau})^2 
- \frac{1}{2\bg_s^2}(S^2 - \frac{M(\tau)\bg_s}{3} S)^2 ].
\end{equation}
Using the variables
\begin{equation}
\xi (t) = \frac{3}{M (\tau) g_s} S (\tau) e^{Q\tau},
\qquad t \equiv Q \tau,
\end{equation}
it is straightforward to see that the equation of motion becomes
\begin{equation} \label{Rhoeq2}
\frac{d^2 \xi}{dt^2}
- 2 \frac{d \xi}{dt} \left(1 - \frac{1}{2} \frac{B'(t)}{B(t)} \right)
+ \xi \left(1 -\frac{B'(t)}{B(t)} - \frac{B'(t)^2}{4 B(t)^2}
             + \frac{B''(t)}{2 B(t)}
       \right)
+ B(t) (1-2\xi)(\xi-\xi^2)
= 0,
\end{equation}
where
\begin{equation}\label{defB}
B(t) \equiv \left( \frac{M (t)}{3Q} \right)^2.
\end{equation}
This is identical in form to \eqref{Rhoeq}.
Therefore the time dependence of the radius of the fuzzy
spheres $S(\tau)$ is identical to that of the
radius of giant gravitons in the previous section.

The fact that the dynamics of these fuzzy spheres is in
exact agreement with the dynamics of giant gravitons is
an important check that our guess for the Matrix string
action---in particular the way the various factors of
$e^{Q\tau}$ appear in (\ref{ppmatrixstring})---is indeed correct.

\section{Fluctuations}

In the previous sections we found two simple exact solutions
of the equations of motion for giant gravitons, or equivalently 
fuzzy spheres, in the irreducible representation of $SU(2)$ at 
large N. The first solution appears when $\mu(\tau)$ is a constant,
and has a sphere which contracts exponentially in time. In
the Matrix String Theory definitions of fields, the classical size
of the fuzzy sphere is given by
\ben \label{mconstantsol}
S(\tau) = \frac{Mg_s}{3}\xi_0~e^{-Q\tau},
\een
where
\ben \label{chizero}
\xi_0 = \frac{3}{4} + {\sqrt{\frac{1}{16}-\frac{1}{B}}},
\een
and $B$ is given by (\ref{defB}). The second solution appears when
$\mu (\tau)$ or equivalently $M(\tau)$ has the special time dependence
\ben \label{taudepmass}
M(\tau) = M_0~e^{Q\tau}.
\een
This is the case where the supersymmetry of the action is broken only by
boundary conditions. In this case there is a specific solution where
the size of the sphere is {\em constant\/} in time, given by
\ben \label{sconstantsol}
S_0 = \frac{M_0 g_s}{3}.
\een
To examine the physics of these time-dependent solution we need to determine
the dynamics of small fluctuations around these solutions. We have not
performed a complete analysis of these fluctuations. However the dynamics
of a special class of fluctuations can be found from what we have already
done. These are fluctuations of the form
\ben \label{lzerofluc}
X^a (\sigma,\tau) = [ S_0 (\tau) + \varphi (\sigma,\tau) ] J^a,
\een
with all other fields set to zero. In the language of giant gravitons
this means that we are looking at fluctuations which are spherically
symmetric on the worldvolume. In this section we will determine the
spectrum around the two special solutions (\ref{mconstantsol}) and
(\ref{sconstantsol}).

\subsection{$M = \text{constant}$: 
Fluctuations around the decaying sphere} \label{sec:decayfluc}

The equations of motion of quadratic fluctuations around
(\ref{mconstantsol}) can be written down by considering perturbations
of the equations (\ref{Rhoeq2}) supplemented by a $\sigma$-derivative
term. These become simple in terms of the variables
\ben \label{deftvphi}
{\tilde{\varphi}} = \frac{3}{Mg_s}~e^{Q\tau}~\varphi,
\een
and the equation for $\tvphi$ is
\ben
\frac{d^2 \tvphi}{dt^2}- \frac{1}{Q^2}\frac{d^2 \tvphi}{d\sigma^2}
- 2 \frac{d\tvphi}{dt}+\nu^2 \tvphi = 0,
\een
where we have used $t = Q\tau$, and
\ben
\nu^2 = A\xi_0 (4\xi_0 - 3).
\een
Note that (\ref{chizero}) implies that $\nu^2 > 0$. The modes of the form
\ben
\tvphi \sim e^{-i\omega t + i k Q \sigma},
\een
have dispersion relation
\ben
\omega = i \pm {\sqrt{k^2+\nu^2 - 4}}.
\een
The imaginary part comes from the anti-damping term in the equation
(\ref{Rhoeq2}). The quantity in the square root may be calculated to be
\ben
\beta^2 = k^2 + \nu^2 -4 =\frac{1}{4}[4k^2 + 12 B {\sqrt{\frac{1}{16}-\frac{1}{B}}}
+ 16 B - 24].
\een
This is explicitly positive since our solution holds only for $B > 8$.
The imaginary part of $\omega$, however means that the solutions for
$\tvphi$ {\em grow with time\/}. This growth is exactly compensated by
the fact that $\tvphi$ defined in (\ref{deftvphi}) 
already contains a factor of $e^{t}$, so that the fluctuations of
the radius itself, $\varphi$ behaves as
\ben
\varphi \sim e^{ik\sigma}~e^{\pm i \frac{\beta}{2}},
\een
which are standard positive and negative frequency oscillator modes.

This seems to suggest that even though the background fuzzy spheres
shrink to zero size at late times, the non-abelian excitations
consisting of spherically symmetric fluctuations of the size of these
spheres have standard oscillatory behavior, corresponding to usual
particles. However, the non-linear terms involving $\varphi$ become large at late times. In fact it is easy to see that the terms which are of order $\varphi^3$ come with an overall coefficient of $e^{Q\tau}$ while those which are $O(\varphi^4)$ have an overall coefficient of $e^{2Q\tau}$. Therefore at late times these terms will constrain $\varphi$ to vanish. Therefore we conclude at late times, not only do the fuzzy spheres shrink to zero size, but the amplitudes of  breathing modes vanish as well. 

\subsection{$M(\tau) \sim e^{Q\tau}$: 
Fluctuations around the static spheres}

The situation is quite different for the case of time dependent $M$ given by (\ref{taudepmass}). In this case it is easier to analyze the situation from
the original action (\ref{faction}) restricted to (\ref{lzerofluc}). The
quadratic action for $\varphi$ may be easily seen to be
\ben \label{radfluct}
S = \int d\tau d\sigma 
[ \frac{1}{2} (\partial_\tau \varphi)^2 - 
\frac{1}{2} (\partial_\sigma \varphi)^2 - 
\frac{1}{2} (\frac{M_0}{3})^2 e^{2Q\tau} \varphi^2 ].
\een
This is the action of a field with time-dependent mass in two
dimensions. By a standard conformal transformation, this is
equivalent to the action of a standard massive field in the Milne
quadrant.

The physics for such a field is quite well known \cite{Birrell:1982ix,
Strominger:2002pc}. In Milne time, the in and out vacua are
inequivalent. This leads to particle production. If we start out with
vacuum at the big bang one ends up with a thermal state at late
times. For our problem, the reverse situation is more interesting. If
we demand that the state at late times is the vacuum of the $\varphi$
particles, we have to start out with a thermal state at the big bang.

\subsection{General fluctuations for an irreducible sphere}

The general problem of small fluctuations can be treated as
in~\cite{Das:2003yq}.  We consider this not just to reanalyze the
results of the previous subsections in more generality, but also to
address the question---especially considering the growth of the
fluctutations---of whether the truncation of the problem
to~\eqref{faction} is physically reasonable.  For the standard
\ppwave---$Q=0$ and constant $\mu$---fuzzy spheres minimize the
energy.  This is also true for the static spheres.
For the general problem however, the spheres are not static, and so the
corresponding solution to the equations of motion
sets neither the potential energy nor the kinetic energy to zero.
We would therefore like to at least check that the solutions
we have been discussing are (perturbatively)
stable.  That is, we check that there are no
negative modes in the spectrum of small oscillations about the solution.

Upon expanding the fields in terms of matrix
spherical harmonics,%
\footnote{We will attempt to be brief.  Further details on matrix scalar and
vector spherical harmonics can be found in~\cite{Das:2003yq}.
These details include
the range of summation, and reality properties of the fluctuating fields
in the expansion.}
\begin{equation}
\begin{gathered}
X^a = S(\tau) J^a + \sum_{j,l,m} x_{jlm} Y^a_{jlm},\\
\begin{aligned}
X^{a''} &= \sum_{lm} x^{a''}_{lm} Y_{lm}, & \qquad
X^8 &= \sum_{lm} x^8_{lm} Y_{lm}, & \qquad
A_\alpha &= \sum_{lm} a_{\alpha l m} Y_{lm},
\end{aligned}
\end{gathered}
\end{equation}
one quickly finds that the dynamics of the fluctuations $x^{a''}_{lm}$
are governed by the action
\begin{equation}
S = \int d \tau d\sigma \left\{
\abs{\dot{x}^{a''}_{l m}}^2
- \abs{x^{\prime a''}_{l m}}^2
- \left[ \frac{l(l+1) S(\tau)^2}{\bg_s^2}
  + \left(\frac{M(\tau)}{6}\right)^2\right] \abs{x^{a''}_{l m}}^2
\right\}.
\end{equation}

Note that these are positive ``mass'' fluctuations.  For the static fuzzy
spheres, the mass is precisely that of~\cite{Das:2003yq}, dressed with
the time-dependence of the Milne universe.

Similarly, the fluctuations $x_{l\pm 1, l, m}$ are described by
\begin{align}
S &= \int d \tau d\sigma \left\{
\abs{\dot{x}_{l+1,l,m}}^2
- \abs{x^\prime_{l+1,l,m}}^2
- \frac{1}{2 \bg_s^2} \left(\frac{M(\tau) \bg_s}{3} + S(\tau) l \right)^2
\abs{x_{l+1,l m}}^2
\right\},
\\
S &= \int d \tau d\sigma \left\{
\abs{\dot{x}_{l-1,l,m}}^2
- \abs{x^\prime_{l-1,l,m}}^2
- \frac{1}{2 \bg_s^2} \left(\frac{M(\tau) \bg_s}{3}
    - S(\tau) (l+1) \right)^2
\abs{x_{l-1,l,m}}^2
\right\}.
\end{align}
These are again positive, and, for the static sphere, are the Milne
universe analogs of the results of~\cite{Das:2003yq}.
The radial fluctuation of $S(\tau)$ itself is encoded in $x_{010}$
(\cf~\eqref{radfluct}).

The remaining fluctuations are coupled in a rather complicated way.
The analysis is performed, as in~\cite{Das:2003yq}, by
introducing auxiliary scalar fields $\phi_{lm}$
which facilitate integrating out the
gauge field.
The result is independent of the gauge degrees of freedom $x_{llm}$.
The case of $l=0$ is special, and we temporarily defer that
interesting discussion.  For $l\neq 0$, one finds the action
\begin{multline}
S = \int d\tau d\sigma \left\{
\abs{\dot{x}^8_{lm}}^2 - \abs{x^{\prime8}_{lm}}^2
 + \frac{1}{S(\tau)^2 l(l+1)} \left[ 
     \abs{\tfrac{\p}{\p \tau} \left(\omega_l \bg_s \phi_{lm}\right)}^2
     - \omega_l^2 \bg_s^2 \abs{\phi'_{lm}}^2
\right] 
\right. \\ \left.
- \left[ \frac{M(\tau)^2}{9} + \frac{S(\tau)^2}{\bg_s^2} l(l+1)\right]
    \abs{X^8_{l m}}^2
+ \frac{M(\tau)}{3} \omega_l (\phi^*_{lm} X^8_{lm} + \phi_{lm} X^{8*}_{lm})
\right\}.
\end{multline}
Here $\omega_l$ is the so-far unspecified and arbitrary
(possibly time-dependent)
normalization of $\phi_{lm}$.  Note that, unlike the
\ppwave\ considered in~\cite{Das:2003yq}, the problem appears to be
nontrivial to untangle.  However, it is clear that the mass matrix is
positive-definite, as is the kinetic term, so this suggests that the
solution is stable.  That said, the friction term in the equation of
motion for $\phi$ could result in a growth of fluctuations.
Finally, let us make our now-familiar comment that,
for the static spheres,
upon choosing
the normalization $\omega_l = \frac{M(\tau)}{3} \sqrt{l(l+1)}$
the masses again precisely reproduce the
Milne universe analogues of the masses of~\cite{Das:2003yq}.

For $l=0$, the result is quite simple,
\begin{equation}
S = \int d\tau d\sigma \left\{
\abs{\dot{x}^8_{00}}^2 - \abs{x^{\prime8}_{00}}^2
- \abs{\frac{M(\tau)}{3} X^8_{00} - \omega_l \phi_{00}}^2
\right\}, \qquad \omega_l \bg_s \phi_{00} = \text{const}.
\end{equation}
Thus, for static fuzzy spheres, it is natural to choose $\omega_l =
\frac{M(\tau)}{3}$.  This implies $\phi_{00} = \text{const}$, and
allows us to preserve a minimum energy by taking $X^8_{00} = \phi_{00}
= \text{const}$.  These are precisely the solutions
of~\cite{Das:2003yq} which corresponded, upon lifting back up to
11-dimensions, to ``orbiting'' fuzzy spheres.  Their
11-dimensional interpretation
for $Q\neq 0$ is less clear.

Finally, let us make a comment about constant $\mu$ and $M$.  In this
case, $S(\tau)$ falls off exponentially like $e^{-Q \tau}$ at late
times.  Therefore, the fluctuations $x^{a''}$ and $x_{l\pm1,l,m}$
asymptote to fluctuations in terms of a constant mass, as was found
in~$\S$\ref{sec:decayfluc}.
Upon choosing $\omega_l = \bg_s^{-1} \sqrt{l(l+1)}$, one similarly finds
a constant mass matrix for $x^8_{lm}$ and $\phi_{lm}$ ($l \neq 0$)
but a nontrivial (damping!) kinetic term for the $\phi_{lm}$.
The zero mode of $X^8$, however, loses its shift symmetry.

\section{Discussion}

As argued in
\cite{Craps:2005wd}, in the class of backgrounds considered in
this paper,
one would expect non-abelian
configurations of Matrix String Theory to be important at early times,
while at late times, the excitations are dominated by usual strings.
We have shown that \ppwave\ geometries provide a useful laboratory to
investigate the detailed time evolution of a class of non-abelian
configurations. The presence of another length scale in the problem,
{\em viz.\/} the strength of the background flux $\mu$, leads to a natural
class of such non-abelian configuration spheres. The
picture is most transparent for constant $\mu$. When $Mg_s \gg 1$
classical configurations dominate the dynamics. The classical dynamics
is controlled by a single combination of the background flux $\mu$ and
the rate of change of the dilaton $Q$, {\em viz.\/} $B= M/(3Q)$. For large
enough $B$, there is a special class of solutions for which the time
dependence of the radius is a pure exponential. At early times, when
the IIA theory is strongly coupled, the radii are big while they go to
zero exponentially at late times. Generic time dependent solutions
have oscillatory behavior; the radii generically oscillate, but with
amplitudes which go to zero at late times. This means that at late
times only diagonal matrices survive, which lead to usual perturbative
strings. While we expect that this is generically true for {\em all}
non-abelian configurations, we do not have any definitive results for
this.

One interesting feature of fuzzy sphere configurations of usual
\ppwave\ matrix string theory is the fact that the fluctuations around
the classical solutions correspond to fields which do not generically
return to their original values as we go around the compact circle any
number of times. It would be interesting to see how these reduce to
usual periodic strings as the size of the fuzzy spheres go to zero at
late times.

It is clear from our discussion that \ppwave\ solutions with linear
null dilatons may be easily obtained in Type IIB theory, \eg\ for the
Penrose limit of $AdS_5 \times S^5$. The behavior of giant gravitons
in such backgrounds may be investigated along the lines of this
paper. However in this case the bulk geometry has a different
holographic description in terms of the large R-charge limit of $N=4$
Yang-Mills theory and giants can be identified as specific states in
this gauge theory. It would be interesting to investigate the implications
of a linear null dilaton for the Yang-Mills theory.

\acknowledgments

We thank M.~Berkooz, B.~Craps, C.~Gomez, P.~Mukhopadhyay,
A.~Shapere, S.~Trivedi and C.~Zhou
for useful conversations.  This work was supported in part by
Department of Energy contract
\#DE-FG01-00ER45832 and the National Science Foundation grant
No. PHY-0244811. S.R.D. would like to thank Tata Institute of
Fundamental Research for hospitality and the organizers of the
Benasque String Theory Workshop for providing a wonderful atmosphere.

\appendix

\section{Supersymmetry Transformation of the Matrix String Action}
\label{sec:genSUSY}

In this appendix we give a general formula for the supersymmetry
transformation of the action~\eqref{ppmatrixstring}.

Since, for $\mu=0$, supersymmetry is manifest in
flat coordinates, we perform the coordinate
transformation~\cite{Craps:2005wd}
\begin{equation}
\pm \tau + \sigma = \pm \tfrac{1}{Q} \ln( \sqrt{2} Q \xi^\pm ),
\qquad
\xi^\pm = \frac{1}{\sqrt{2}} ( \xi^0 \pm \xi^1 ).
\end{equation}
Then the Matrix String action is
\begin{multline} \label{MST:XI} 
S =  S_{\text{fermion}} + \int d^2\xi \Tr \left\{
- \frac{1}{4} g_s^2 F_{\alpha\beta}^2
- \frac{1}{2} (D_\alpha X^i)^2 
+ \frac{1}{4 g_s^2} \com{X^i}{X^j}^2 
\right. \\ \left.
- \frac{1}{2} \left(\frac{\tM (\xi) }{3}\right)^2 (X^a)^2
-\frac{1}{2} \left(\frac{\tM (\xi) }{6}\right)^2 (X^{a''})^2
-\frac{\tM (\xi) }{6} g_s X^8 \epsilon^{\alpha\beta} 
F_{\alpha\beta}
- i \frac{\tM (\xi)}{3 g_s} \epsilon_{abc} X^a X^b X^c
\right\},
\end{multline}
where
\begin{equation}
\tM(\xi) = \frac{\mu(\xi) l_s^2}{Q R \xi},
\qquad \xi \equiv \sqrt{2 \xi^+ \xi^-} = \tfrac{1}{Q} e^{Q \tau}.
\end{equation}
For simplicity, we have written $\mu(\xi)$ instead of $\mu(\tau(\xi))$.
In contrast to the main text,
we use nonchiral, Majorana SO(8) fermions $\Psi$, and lose manifest
SO(1,1) covariance, so that
\begin{multline}
S_{\text{fermion}} = \int d^2\xi \Tr \left\{ 
i\Psi^\transpose D_0 \Psi
- i\Psi^\transpose \Gamma^9 D_1 \Psi
\right. \\ \left.
 +
\frac{1}{g_s} \Psi^\transpose \Gamma^i \com{X^i}{\Psi}
-i\frac{\tM(\xi)}{4} \Psi^\transpose
      (\Gamma^{123} - \frac{1}{3} \Gamma^{89}) \Psi
\right\}.
\end{multline}

When $\mu$, and therefore $M$, is constant, and $Q=0$, the action has
eight linearly realized supersymmetries parametrized by
$\epsilon_0 = \Gamma^{4567} \epsilon_0 = \Gamma^{12389} \epsilon_0$.%
\footnote{The last expression is the statement that we use the
convention $\Gamma^{123456789} = \one$.}  In overcomplete
generality, one can consider the supersymmetry transformations
\begin{subequations} \label{genMppSUSY}
\begin{gather} \label{genMppSUSYX}
\begin{align} 
\delta X^i &= i \alpha_1(\xi) \Psi^\transpose \Gamma^i \epsilon, &
\delta A_0 &= i \alpha_3(\xi) \Psi^\transpose \epsilon, &
\delta A_1 &= i \alpha_2(\xi) \Psi^\transpose \Gamma^9 \epsilon,
\end{align} \\ \label{genMppSUSYPsi}
\begin{align}
\delta \Psi &= \frac{1}{2} \alpha_4(\xi) F_{01} \Gamma^9 \epsilon
 + \frac{1}{2} \alpha_5(\xi) D_0 X^i \Gamma^i \epsilon
 - \frac{1}{2} \alpha_6(\xi) D_1 X^i \Gamma^i \Gamma^9 \epsilon
 + \frac{i}{4} \alpha_7(\xi) \com{X^i}{X^j} \Gamma^{ij} \epsilon
\\ & \qquad
 + \alpha_8(\xi) \frac{\tM(\xi)}{6} X^a \Gamma^a \Gamma^{123} \epsilon
 - \alpha_9(\xi) \frac{\tM(\xi)}{12} X^{a''} \Gamma^{a''}\Gamma^{123} \epsilon,
\end{align}
\intertext{where} \label{genepsMppSUSY}
\epsilon = \Gamma^{4567} \epsilon_0 = \Gamma^{12389} \epsilon_0.
\end{gather}
\end{subequations}
For $Q=0$, $\tM(\xi)$ constant and $\alpha_\alpha(\xi)=1$, this reduces
to the supersymmetry transformations given in~\cite{Das:2003yq}, and
preserves the action.  In the current context, the variation of the
action is
\begin{subequations}
\begin{equation}
\delta S = \delta S_{\text{t.d.}} + \delta S_{\text{1}}
 + \delta S_{\text{\p}} + \delta S_{\tM},
\end{equation}
where the various pieces of the action are
\begin{multline} \label{SUSY:alpha}
\delta S_{\text{1}} = \int d\xi^0 d\xi^1 \Tr \left\{
i (\alpha_1-\alpha_4) \frac{\tM}{3} F_{01} \epsilon^\transpose \Gamma^8 \Psi
+ i (\alpha_2-\alpha_4) F_{01} \epsilon^\transpose D_1 \Psi
\right. \\ \left.
- i (\alpha_3 - \alpha_4) F_{01} \epsilon^\transpose \Gamma^9 D_0 \Psi
+ \left(\tfrac{\alpha_5+\alpha_6}{2}-\alpha_4\right) \com{F_{01}}{X^i}
     \epsilon^\transpose \Gamma^i \Gamma^9 \Psi
+ i \tfrac{\alpha_5-\alpha_6}{2} \anti{D_0}{D_1}
     \epsilon^\transpose \Gamma^i \Gamma^9 \Psi
\right. \\ \left.
+ \tfrac{i}{3} (\alpha_8-\alpha_5) \tM D_0 X^a
     \epsilon^\transpose \Gamma^a \Gamma^{123} \Psi
+ \tfrac{i}{6} (\alpha_9 - \alpha_5) \tM D_0 X^{a''}
     \epsilon^\transpose \Gamma^{a''} \Gamma^{123} \Psi
\right. \\ \left.
- \tfrac{i}{3} (\alpha_3 - \alpha_5) \tM D_0 X^8
     \epsilon^\transpose \Gamma^9 \Psi
+ i (\alpha_1-\alpha_8) \bigl(\tfrac{\tM}{3}\bigr)^2 X^a
      \epsilon^\transpose \Gamma^a \Psi
\right. \\ \left.
+ i (\alpha_1-\alpha_9) \bigl(\tfrac{\tM}{6}\bigr)^2 X^{a''}
      \epsilon^\transpose \Gamma^{a''} \Psi
+ \tfrac{i}{3} (\alpha_6-\alpha_8) \tM D_1 X^a
     \epsilon^\transpose \Gamma^a \Gamma^{123} \Gamma^9 \Psi
\right. \\ \left.
+ \tfrac{i}{6} (\alpha_6 - \alpha_9) \tM D_1 X^{a''}
     \epsilon^\transpose \Gamma^{a''} \Gamma^{123} \Gamma^9 \Psi
- \tfrac{i}{3} (\alpha_6 - \alpha_2) \tM D_1 X^8
     \epsilon^\transpose \Psi
\right. \\ \left.
-i(\alpha_1-\alpha_5) D_0 X^i \epsilon^\transpose \Gamma^i D_0 \Psi
+i(\alpha_1-\alpha_6) D_1 X^i \epsilon^\transpose \Gamma^i D_1 \Psi
\right. \\ \left.
+ (\alpha_5-\alpha_2) D_0 X^i \epsilon^\transpose \com{X^i}{\Psi}
+ (\alpha_5-\alpha_7) D_0 X^i \epsilon^\transpose \Gamma^{ij} \com{X^j}{\Psi}
\right. \\ \left.
+ (\alpha_3-\alpha_6) D_1 X^i \epsilon^\transpose \Gamma^9 \com{X^i}{\Psi}
+ (\alpha_7-\alpha_6) D_1 X^i
      \epsilon^\transpose \Gamma^{ij} \Gamma^9 \com{X^j}{\Psi}
\right. \\ \left.
-\tfrac{\tM}{3} (\alpha_8 + \tfrac{1}{2} \alpha_7 - \tfrac{3}{2} \alpha_1)
  \com{X^a}{X^b} \epsilon^\transpose \Gamma^{ab} \Gamma^{123} \Psi
+\tfrac{\tM}{3} (\alpha_8 - \tfrac{1}{2} \alpha_9 - \tfrac{1}{2} \alpha_7)
  \com{X^a}{X^{b''}} \epsilon^\transpose \Gamma^{ab''} \Gamma^{123} \Psi
\right. \\ \left.
+\tfrac{\tM}{3} (\alpha_8 - \alpha_7)
  \com{X^a}{X^8} \epsilon^\transpose \Gamma^{a8} \Gamma^{123} \Psi
+\tfrac{\tM}{6} (\alpha_9 - \alpha_7)
  \com{X^{a''}}{X^{b''}} \epsilon^\transpose \Gamma^{a''b''} \Gamma^{123} \Psi
\right. \\ \left.
+\tfrac{\tM}{6} (\alpha_9 - \alpha_7)
  \com{X^{a''}}{X^8} \epsilon^\transpose \Gamma^{a''8} \Gamma^{123} \Psi
+ i (\alpha_1-\alpha_7) \com{\com{X^i}{X^j}}{X^j}
   \epsilon^\transpose \Gamma^i \Psi
\right\};
\end{multline}
\begin{multline} \label{SUSY:pa}
\delta S_{\text{\p}} = \int d\xi^0 d\xi^1 \Tr \left\{
i \p_1 \alpha_2 F_{01} \epsilon^\transpose \Psi
- i \p_0 \alpha_3 F_{01} \epsilon^\transpose \Gamma^9 \Psi
- i \p_0 \alpha_6 D_1 X^i \epsilon^\transpose \Gamma^i \Gamma^9 \Psi
\right. \\ \left.
+ i \p_1 \alpha_5 D_0 X^i \epsilon^\transpose \Gamma^i \Gamma^9 \Psi
- i \p_0 \alpha_1 D_0 X^i \epsilon^\transpose \Gamma^i \Psi
+ i \p_1 \alpha_1 D_1 X^i \epsilon^\transpose \Gamma^i \Psi
\right. \\ \left.
-\tfrac{1}{2} \p_0 \alpha_7 \com{X^i}{X^j}
     \epsilon^\transpose \Gamma^{ij} \Psi
+\tfrac{1}{2} \p_1 \alpha_7 \com{X^i}{X^j}
     \epsilon^\transpose \Gamma^{ij} \Gamma^9 \Psi
\right\};
\end{multline}
\begin{multline} \label{SUSY:tM}
\delta S_{\tM} = \int d\xi^0 d\xi^1 \Tr \left\{
\tfrac{i}{6} \p_0 (\alpha_9 \tM) X^{a''} 
      \epsilon^\transpose \Gamma^{a''} \Gamma^{123} \Psi
- \tfrac{i}{3} \alpha_3 \p_0 \tM X^8 
      \epsilon^\transpose \Gamma^9 \Psi
\right. \\ \left.
+ \tfrac{i}{3} \p_0 (\alpha_8 \tM) X^a 
      \epsilon^\transpose \Gamma^a \Gamma^{123} \Psi
- \tfrac{i}{3} \p_1 (\alpha_8 \tM) X^a 
      \epsilon^\transpose \Gamma^a \Gamma^{123} \Gamma^9 \Psi
\right. \\ \left.
- \tfrac{i}{6} \p_1 (\alpha_9 \tM) X^{a''} 
      \epsilon^\transpose \Gamma^{a''} \Gamma^{123} \Gamma^9 \Psi
+ \tfrac{i}{3} \alpha_2 \p_1 \tM X^8 
      \epsilon^\transpose \Psi
\right\};
\end{multline}
and the total derivatives are
\begin{multline} \label{SUSY:td}
\delta S_{\text{t.d.}} = \int d^2\xi \Tr \left\{
\p_0 \left[
\tfrac{i}{2} \alpha_6 D_1 X^i \epsilon^\transpose \Gamma^i \Gamma^9 \Psi
-\tfrac{i}{6} \alpha_8 \tM X^a \epsilon^\transpose \Gamma^a \Gamma^{123} \Psi
-\tfrac{i}{12} \alpha_9 \tM X^{a''} \epsilon^\transpose \Gamma^{a''}
    \Gamma^{123} \Psi
\right. \right. \\ \left. \left.
-\tfrac{i}{3} \alpha_3 \tM X^8 \epsilon^\transpose \Gamma^8 \Gamma^{123} \Psi
+\tfrac{1}{4} \alpha_7 \com{X^i}{X^j} \epsilon^\transpose \Gamma^{ij} \Psi
- \tfrac{i}{2} \alpha_4 F_{01} \epsilon^\transpose \Gamma^9 \Psi
- \tfrac{i}{2} \alpha_5 D_0 X^i \epsilon^\transpose \Gamma^i \Psi
\right]
\right. \\ \left.
+ \p_1 \left[ 
\tfrac{i}{2} \alpha_5 D_0 X^i \epsilon^\transpose \Gamma^i \Gamma^9 \Psi
+\tfrac{i}{6} \alpha_8 \tM X^a \epsilon^\transpose \Gamma^a \Gamma^{123} 
   \Gamma^9 \Psi
+\tfrac{i}{12} \alpha_9 \tM X^{a''} \epsilon^\transpose \Gamma^{a''}
    \Gamma^{123} \Gamma^9 \Psi
\right. \right. \\ \left. \left.
+\tfrac{i}{3} \alpha_2 \tM X^8 \epsilon^\transpose \Gamma^8 \Gamma^{123}
    \Gamma^9 \Psi
-\tfrac{1}{4} \alpha_7 \com{X^i}{X^j} \epsilon^\transpose \Gamma^{ij} 
    \Gamma^9 \Psi
+ \tfrac{i}{2} \alpha_4 F_{01} \epsilon^\transpose \Psi
+ \tfrac{i}{2} \alpha_6 D_1 X^i \epsilon^\transpose \Gamma^i \Psi
\right]
\right\}.
\end{multline}
\end{subequations}
The coefficients of the various combinations of fields in
Eq.~\eqref{SUSY:alpha} vanish only if the $\alpha$'s are equal.
Eq.~\eqref{SUSY:pa} then yields that the $\alpha$'s should be
constant, whereupon Eq.~\eqref{SUSY:tM} then implies that $\tM$ should
be constant as well.  That is, there are supersymmetry currents when
the bulk part of the integrand vanishes.  This happens (only, by choice
of normalization) for $\alpha_1 = \alpha_2 =
\dots = \alpha_9 = 1$ and for $\xi$-independent $\tM$.  This clearly
occurs for the maximally supersymmetric \ppwave, but also occurs by
taking $\mu(\xi) \propto
\xi \propto e^{Q \tau}$.
Of course, as in~\cite{Craps:2005wd}, even in that case,
the total derivatives~\eqref{SUSY:td}
contribute because of the nontrivial topology of the Milne universe.

\section{Isometries} \label{sec:killing}

It is interesting to work out the Killing vectors of the IIA
background~\eqref{maxdil2a}.  We present those here---see also~\cite{prt}---%
and mimic the notation of~\cite{fp}.

The obvious Killing vectors are the null isometry,
\begin{equation}
e_- = -\frac{\p}{\p x^-},
\end{equation}
and the rotational Killing vectors,
\begin{align}
J_{ab} &= x^a \frac{\p}{\p x^b} - x^b \frac{\p}{\p x^a}, &
J_{a'' b''} &= x^{a''} \frac{\p}{\p x^{b''}} - x^{b''} \frac{\p}{\p x^{a''}}.
\end{align}
The less obvious Killing vectors are (no sum)
\begin{equation}
\begin{aligned}
e_i &= k^{(1)}_i(x^+) \frac{\p}{\p x^i}
  - x^i \p_+ k^{(1)}_i(x^+) \frac{\p}{\p x^-}, \\
e^*_i &= k^{(2)}_i(x^+) \frac{\p}{\p x^i}
   - x^i \p_+ k^{(2)}_i(x^+) \frac{\p}{\p x^-},
\end{aligned}
\end{equation}
where $k^{(1)}_i$ and $k^{(2)}_i$ are the two solutions to the second order
differential equation,
\begin{equation} \label{Kill:diffeq}
0 = 
\begin{cases}
\p_+^2 k_i(x^+) + \left(\frac{\mu(x^+)}{3}\right)^2 k_i(x^+), & i=1,2,3,8, \\
\p_+^2 k_i(x^+) + \left(\frac{\mu(x^+)}{6}\right)^2 k_i(x^+), & i=4,5,6,7.
\end{cases}
\end{equation}
For constant $\mu$, these are precisely the Killing vectors previously given
in the literature (\eg~\cite{fp}) and for the case $\mu(x^+) = \mu e^{Q x^+}$,
one has, in terms of Bessel and Neumann functions,
\begin{align}
k_i^{(1)}(x^+) &= J_0(\hat{\xi}_i), &
k_i^{(2)}(x^+) &= N_0(\hat{\xi}_i), &
\hat{\xi}_i = \begin{cases} \frac{\mu}{3 Q} \xi, & i=1,2,3,8, \\
                            \frac{\mu}{6 Q} \xi, & i=4,5,6,7. \end{cases}
\end{align}

The isometry algebra is therefore 8 copies of the Heisenberg algebra,
with a common central element, proportional to $e_-$ with
a normalization that is given by
the (constant!) Wronskian of the two solutions to the differential
equation~\eqref{Kill:diffeq}.


\begin{thebibliography}{99}


\bibitem{Craps:2005wd}
  B.~Craps, S.~Sethi and E.~P.~Verlinde,
  \ct{A matrix big bang}
  \hepth{0506180}.

\bibitem{Motl:1997th}
  L.~Motl,
  \ct{Proposals on nonperturbative superstring interactions}
  \hepth{9701025}.

\bibitem{Banks:1996my}
  T.~Banks and N.~Seiberg,
  \ct{Strings from matrices}
  \npb{497}{1997}{41};
  \phepth{9702187}.

\bibitem{Dijkgraaf:1997vv}
  R.~Dijkgraaf, E.~P.~Verlinde and H.~L.~Verlinde,
  \ct{Matrix string theory}
  \npb{500}{1997}{43};
  \phepth{9703030}.

\bibitem{hs}
G.~T.~Horowitz and A.~R.~Steif,
\ct{Space-time Singularities in String Theory}
\citeprl{64}{1990}{260}.

\bibitem{bhkn}
V.~Balasubramanian, S.~F.~Hassan, E.~Keski-Vakkuri and A.~Naqvi,
\ct{A space-time orbifold: A toy model for a cosmological singularity}
\citeprd{67}{2003}{026003};
\phepth{0202187}.

\bibitem{cc}
L.~Cornalba and M.~S.~Costa,
\ct{A new cosmological scenario in string theory}
\citeprd{66}{2002}{066001};
\phepth{0203031}.

\bibitem{s}
J.~Simon,
\ct{The geometry of null rotation identifications}
\jhep{06}{2002}{001};
\phepth{0203201}.

\bibitem{lms}
H.~Liu, G.~Moore and N.~Seiberg,
\ct{String in a time-dependent orbifold}
\jhep{06}{2002}{045};
\phepth{0204168}.

\bibitem{lms2}
H.~Liu, G.~Moore and N.~Seiberg,
\ct{String in time-dependent orbifolds}
\jhep{10}{2002}{031};
\phepth{0205288}.

\bibitem{tt}
A.~J.~Tolley and N.~Turok,
\ct{Quantum Fields in a Big Crunch/Big Bang Spacetime}
\citeprd{66}{2002}{106005};
\phepth{0204091}.

\bibitem{ckr}
B.~Craps, D.~Kutasov and G.~Rajesh,
\ct{String propagation in the presence of cosmological singularities}
\jhep{06}{2002}{053};
\phepth{0205101}.

\bibitem{fm}
M.~Fabinger and J.~McGreevy,
\ct{On smooth time-depedent orbifolds and null singularities}
\jhep{06}{2003}{042};
\phepth{0206196}.

\bibitem{d}
J.~R.~David,
\ct{Plane waves with weak singularities}
\jhep{11}{2003}{064};
\phepth{0303013}.

\bibitem{cm}
B.~C.~Da Cunha and E.~J.~Martinec,
\ct{Closed string tachyon condensation and worldsheet inflation}
\citeprd{68}{2003}{063502}
\phepth{0303087}.

\bibitem{r}
J.~G.~Russo,
\ct{Cosmological string models from Milne spaces and SL(2,\ZZ) orbifold}
\mpla{19}{2004}{421--432};
\phepth{0305032}.

\bibitem{grs}
A.~Giveon, E.~Rabinovici and A. Sever,
\ct{Strings in singular time-dependent backgrounds}
\citejournal{51}{2003}{805--823}{Fortsch.\ Phys.\ };
\phepth{0305137}.

\bibitem{pb}
B.~Pioline and M.~Berkooz,
\ct{Strings in an electric field, and the Milne universe}
\citejournal{11}{2003}{007}{J.\ Cosmol.\ Astropart.\ Phys.\ };
\phepth{0307280}.

\bibitem{dp}
B.~Durin and B.~Pioline,
\ct{Closed strings in Misner space: A toy model for a big bounce?}
LPTHE-P05-03,
\hepth{0501145}.

\bibitem{bdpr}
M.~Berkooz, B.~Durin, B.~Pioline and D.~Reichmann,
\ct{Closed strings in Misner space: Stringy fuzziness with a twist}
\citejournal{10}{2004}{002}{J.\ Cosmol.\ Astropart.\ Phys.\ };
\hepth{0407216}.

\bibitem{Das:2004aq}
  S.~R.~Das and J.~L.~Karczmarek,
  \ct{Spacelike boundaries from the c = 1 matrix model}
  \citeprd{71}{2005}{086006};
  \phepth{0412093}.

\bibitem{Das:2005jp}
  S.~R.~Das,
  \ct{Non-trivial 2d space-times from matrices}
  \hepth{0503002}.


\bibitem{McGreevy:2005ci}
  J.~McGreevy and E.~Silverstein,
  \ct{The tachyon at the end of the universe}
  \hepth{0506130}.

\bibitem{Myers:1999ps}
  R.~C.~Myers,
  \ct{Dielectric-branes}
  \jhep{12}{1999}{022};
  \phepth{9910053}.

\bibitem{TV}
W.~Taylor and M.~Van~Raamsdonk,
\ct{Multiple Dp-branes in Weak Background Fields}
\npb{573}{2000}{703--734};
\phepth{9910052}.

\bibitem{Berenstein:2002jq}
  D.~Berenstein, J.~M.~Maldacena and H.~Nastase,
  \ct{Strings in flat space and pp waves from N = 4 super Yang Mills}
  \jhep{04}{2002}{013};
  \phepth{0202021}.

\bibitem{clp2} M. Cveti\v{c}, H. L\"{u} and C. N. Pope,
\ct{M-theory, PP-Waves, Penrose Limits and Supernumerary Supersymmetries}
\npb{644}{2002}{65--84};
\phepth{0203229}.

\bibitem{ni} N. Iizuka, \ct{Supergravity, Supermembrane and M(atrix)
model on PP-Waves}
\citeprd{68}{2003}{126002};
\phepth{0211138}.

\bibitem{Dasgupta:2002hx}
  K.~Dasgupta, M.~M.~Sheikh-Jabbari and M.~Van Raamsdonk,
  \ct{Matrix perturbation theory for M-theory on a PP-wave}
  \jhep{05}{2002}{056};
  \phepth{0205185}.

\bibitem{Dasgupta:2002ru}
  K.~Dasgupta, M.~M.~Sheikh-Jabbari and M.~Van Raamsdonk,
  \ct{Protected multiplets of M-theory on a plane wave}
  \jhep{09}{2002}{021};
  \phepth{0207050}.



\bibitem{Kim:2002if}

  N. W.~Kim and J.~Plefka,
  \ct{On the spectrum of pp-wave matrix theory}
  \npb{643}{2002}{31};
  \phepth{0207034}.



\bibitem{Kim:2002zg}
  N.~Kim and J.~H.~Park,
  \ct{Superalgebra for M-theory on a pp-wave}
  \citeprd{66}{2002}{106007};
  \phepth{0207061}.

\bibitem{Michelson:2004fh}
  J.~Michelson,
  \ct{Matrix theory of pp waves}
  in \bt{Quantum Theory and Symmetries: Proceedings of the 3rd International
  Symposium} (P.~Argyres \etal\ ed.) (World Scientific: New Jersey, 2004);
  \phepth{0401050}.

\bibitem{jhp} J. H. Park,
\ct{Supersymmetric objects in the M-theory on a pp-wave}
\jhep{10}{2002}{032};
\phepth{0208161}.

\bibitem{Michelson:2002wa}

  J.~Michelson,
  \ct{(Twisted) toroidal compactification of pp-waves}
  \citeprd{66}{2002}{066002};
  \phepth{0203140}.

\bibitem{Sugiyama:2002tf}
  K.~Sugiyama and K.~Yoshida,
  \ct{Type IIA string and matrix string on pp-wave}
  \npb{644}{2002}{128};
  \phepth{0208029}.

\bibitem{Hyun:2002wu}
  S. J.~Hyun and H. J.~Shin,
  \ct{N = (4,4) type IIA string theory on pp-wave background}
  \jhep{10}{2002}{070};
  \phepth{0208074}.

\bibitem{Hyun:2002wp}
  S. J.~Hyun and H. J.~Shin,
  \ct{Solvable N = (4,4) type IIa string theory in plane-wave background and
  D-branes}
  \npb{654}{2003}{114};
  \phepth{0210158}.

\bibitem{Gopakumar:2002dq}
  R.~Gopakumar,
  \ct{String interactions in PP-waves}
  \citeprl{89}{2002}{171601};
  \phepth{0205174}.

\bibitem{Das:2003yq}
  S.~R.~Das, J.~Michelson and A.~D.~Shapere,
  \ct{Fuzzy spheres in pp-wave matrix string theory}
  \citeprd{70}{2004}{026004};
  \phepth{0306270}.

\bibitem{gb} G. Bonelli,
\ct{Matrix Strings in pp-wave backgrounds from deformed Super Yang-Mills Theory}
\jhep{08}{2002}{022};
\phepth{0205213}.

\bibitem{Shin:2003np}
  H.~Shin and K.~Yoshida,
  \ct{One-loop flatness of membrane fuzzy sphere interaction in plane-wave
  matrix model}
  \npb{679}{2004}{99};
  \phepth{0309258}.

\bibitem{Chen:2003sm}
  Y.~X.~Chen and J.~Shao,
  \ct{Giant graviton in type IIa pp-wave background}
  \citeprd{69}{2004}{106010};
  \phepth{0310062}.

\bibitem{Furuuchi:2003sy}
  K.~Furuuchi, E.~Schreiber and G.~W.~Semenoff,
  \ct{Five-brane thermodynamics from the matrix model}
  \hepth{0310286}.


\bibitem{Janssen:2004jz}
  B.~Janssen, Y.~Lozano and D.~Rodriguez-Gomez,
  \ct{Giant gravitons in AdS$_3 \times $S$^3 \times $T$^4$ as fuzzy cylinders}
  \npb{711}{2005}{392};
  \phepth{0406148}.

\bibitem{Lee:2004kv}
  H.~K.~Lee, T.~McLoughlin and X.~Wu,
  \ct{Gauge / gravity duality for interactions of spherical membranes in
  11-dimensional pp-wave}
CALT-68-2523, UK-04-21,
  \hepth{0409264}.

\bibitem{Lozano:2005kf}

  Y.~Lozano and D.~Rodriguez-Gomez,
  \ct{Fuzzy 5-spheres and pp-wave Matrix actions}
FFUOV-05/02,
  \hepth{0505073}.


\bibitem{Alvarez:1997fy}
  E.~Alvarez and P.~Meessen,
  \ct{Newtonian M(atrix) cosmology}
  \plb{426}{1998}{282};
  \phepth{9712136}.

\bibitem{Freedman:2004xg}
  D.~Z.~Freedman, G.~W.~Gibbons and M.~Schnabl,
  \ct{Matrix cosmology}
  \citejournal{743}{2005}{286}{AIP Conf.\ Proc.\ };
  \phepth{0411119}.

\bibitem{Li:2005sz}
  M.~Li,
  \ct{A class of cosmological matrix models}
  \hepth{0506260}.

\bibitem{Li:2005ti}
  M.~Li and W.~Song,
  \ct{Shock waves and cosmological matrix models}
  \hepth{0507185}.

\bibitem{prt}
G.~Papadopoulos, J.~G.~Russo and A.~A.~Tseytlin,
\ct{Solvable model of strings in a time-dependent plane-wave background}
\cqg{20}{2003}{969-1016};
\phepth{0211289}.

\bibitem{bopt}
M.~Blau, M~O'Loughlin, G.~Papadopoulos and A.~A.~Tseytlin,
\ct{Solvable models of strings in homogeneous plane wave backgrounds}
\npb{673}{2003}{57--97};
\phepth{0304198}.

\bibitem{clp} M. Cveti\v{c}, H. L\"{u} and C. N. Pope,
\ct{Penrose Limits, PP-Waves and Deformed M2-branes}
\citeprd{69}{2004}{046003};
\phepth{0203082}.


\bibitem{fp} J. Figueroa-O'Farrill and
G. Papadopoulos,
\ct{Homogeneous Fluxes, Branes and Maximal Supersymmetry}
\jhep{08}{2001}{036};
\phepth{0105308}.

\bibitem{ghpp} J. P. Gauntlett and C. M. Hull,
\ct{pp-waves in 11-dimensions with extra supersymmetry}
\jhep{06}{2002}{013};
\phepth{0203255}.

\bibitem{rt} J.~G.~Russo and A.~A.~Tseytlin,
\ct{On solvable models of type IIB superstring in NS-NS and R-R plane
wave backgrounds}
\jhep{04}{2002}{021};
\phepth{0202179}.

\bibitem{gms} J.~Gomis, L.~Motl and Andrew Strominger,
\ct{PP-Wave / CFT$_2$ Duality}
\jhep{11}{2002}{016};
\phepth{0206166}.


\bibitem{jm26}
J. Michelson,
\ct{A pp-wave with 26 Supercharges}
\cqg{19}{2002}{5935--5949};
\phepth{0206204}.

\bibitem{seibergsen}
N. Seiberg,
\ct{Why is the Matrix Model Correct?}
\citeprl{79}{1997}{3577--3580};
\phepth{9710009}.

\bibitem{senseiberg}
A.~Sen,
\ct{D0 Branes on T$^n$ and Matrix Theory}
\citejournal{2}{1998}{51--59}{Adv.\ Theor.\ Math.\ Phys.\ };
\phepth{9709220}.

\bibitem{preTV}
W.~I.~Taylor and M.~Van~Raamsdonk,
\ct{Multiple D0-branes in weakly curved backgrounds}
\npb{558}{1999}{63};
\phepth{9904095}.


\bibitem{McGreevy:2000cw}
  J.~McGreevy, L.~Susskind and N.~Toumbas,
  \ct{Invasion of the giant gravitons from anti-de Sitter space}
  \jhep{06}{2000}{008};
  \phepth{0003075}.


\bibitem{Grisaru:2000zn}
  M.~T.~Grisaru, R.~C.~Myers and \O.~Tafjord,
  \ct{SUSY and Goliath}
  \jhep{08}{2000}{040};
  \phepth{0008015}.


\bibitem{Hashimoto:2000zp}
  A.~Hashimoto, S.~Hirano and N.~Itzhaki,
  \ct{Large branes in AdS and their field theory dual}
  \jhep{08}{2000}{051};
  \phepth{0008016}.

\bibitem{Das:2000fu}
  S.~R.~Das, A.~Jevicki and S.~D.~Mathur,
  \ct{Giant gravitons, BPS bounds and noncommutativity}
  \citeprd{63}{2001}{044001};
  \phepth{0008088}.

\bibitem{Das:2000ab}
  S.~R.~Das, S.~P.~Trivedi and S.~Vaidya,
  \ct{Magnetic moments of branes and giant gravitons}
  \jhep{10}{2000}{037};
  \phepth{0008203}.

\bibitem{Das:2000st}
  S.~R.~Das, A.~Jevicki and S.~D.~Mathur,
  \ct{Vibration modes of giant gravitons}
  \citeprd{63}{2001}{024013};
  \phepth{0009019}.

\bibitem{Birrell:1982ix}
N.~D.~Birrell and P.~C.~W.~Davies,
\bt{Quantum Fields in Curved Space}
(Cambridge University Press: Cambridge, 1984).

\bibitem{Strominger:2002pc}
  A.~Strominger,
  \ct{Open string creation by S-branes}
  \hepth{0209090}.

\end{thebibliography}
\end{document}